\documentclass[10pt,lettersize,journal]{IEEEtran}
\usepackage{amsmath,amsfonts, amssymb, bm}
\usepackage{amsthm}
\usepackage{algorithmic}
\usepackage{array}
\usepackage[caption=false,font=normalsize,labelfont=sf,textfont=sf]{subfig}
\usepackage{textcomp}
\usepackage{stfloats}
\usepackage{url}
\usepackage{verbatim}
\usepackage{graphicx}
\usepackage{cite}
\usepackage{pgfplots}
\usepackage{tikz}
\usepackage{etoolbox}
\usetikzlibrary{shapes.misc, positioning, arrows, math, decorations.pathreplacing, patterns}
\usepackage{algorithmic}
\graphicspath{{Figures/}}
\usepackage{textcomp}
\usepackage{xcolor}
\def\BibTeX{{\rm B\kern-.05em{\sc i\kern-.025em b}\kern-.08em
    T\kern-.1667em\lower.7ex\hbox{E}\kern-.125emX}}
\usepackage{balance}
\newcommand{\statevec}[0]{\bm{x}}

\newcommand{\inputvec}[0]{\bm{u}}

\newcommand{\inputmat}[0]{\bm{B}}

\newcommand{\statemat}[0]{\bm{A}}

\newcommand{\noisevec}[0]{\bm{w}}

\newcommand{\sdr}[0]{\mathcal{S}}

\newcommand{\controller}[0]{\mathcal{C}}

\newcommand{\plant}[0]{\mathcal{P}}

\DeclareMathSymbol{\shortminus}{\mathbin}{AMSa}{"39}

\newcommand{\slfrac}[2]{\left.#1\middle/#2\right.}

\definecolor{myred}{RGB}{220,43,25}
\definecolor{mygreen}{RGB}{0,146,64}
\definecolor{myblue}{RGB}{0,143,224}
\definecolor{mygray}{gray}{0.80}
\definecolor{mylightergray}{gray}{0.87}
\definecolor{mylightestgray}{gray}{0.95}
\definecolor{seagreen}{RGB}{46, 139, 87} 
\definecolor{yellowgreen}{RGB}{154, 205, 50} 	
\definecolor{darkyellow}{RGB}{139, 139, 34}

\begin{document}
	
	
	\author{
		\IEEEauthorblockN{Onur Ayan, \textit{Student Member, IEEE}} \\
		\IEEEauthorblockN{Polina Kutsevol} \\
		\IEEEauthorblockN{Hasan Ya\u{g}{\i}z \"{O}zkan} \\

		\IEEEauthorblockN{Wolfgang Kellerer, \textit{Senior Member, IEEE}}
	}
\title{Task-Oriented Scheduling for Networked Control Systems: An Age of Information-Aware Implementation on Software-Defined Radios}

\maketitle

\begin{abstract}
	Networked control systems (NCSs) are feedback control loops that are closed over a communication network. Emerging applications, such as telerobotics, drones and autonomous driving are the most prominent examples of such systems. Regular and timely information sharing between the components of NCSs is essential to fulfill the desired control tasks, as stale information can lead to performance degradation or even physical damage.
	In this work, we consider multiple heterogeneous NCSs that transmit their system state over a shared physical wireless channel towards a gateway node. We conduct a comprehensive experimental study on selected MAC protocols using software-defined radios with state-of-the-art (SotA) solutions that have been designed to increase information freshness and control performance. As a significant improvement over the SotA, we propose a novel contention-free algorithm that is able to outperform the existing solutions by combining their strengths in one protocol. In addition, we propose a new metric called normalized mean squared error that maps the age of information to a dimensionless quantity that captures the expected value of a control system's next transmission. We demonstrate its adoption and effectiveness for wireless resource scheduling in a case study involving multiple inverted pendulums. From our experimental study and results, we observe that value-aware prioritization of the sub-systems contributes to minimizing the negative effects of information staleness on control performance. In particular, as the number of devices increases, the benefit of control-awareness to the quality of control stands out when compared to protocols that focus solely on maximizing information freshness.
\end{abstract}
	
	\section{Introduction and Related Work}
	6G wireless systems are envisioned to be a disruptive generation of cellular networks whose design is tailored to the performance requirements of the supported applications \cite{saad2020vision}.
	In particular, connected robotics and autonomous systems are seen as some of the key driving application domains in 6G wireless systems. Emerging applications such as autonomous cars, autonomous robotics and drone-delivery systems are the most prominent examples of connected robotics. Since such systems rely on regular and timely information sharing between multiple sensors, actuators and controllers, the co-existence of such applications in a network requires a careful end-to-end co-design of communication, control and computing.
	
	From a system theoretic perspective, connected robotics applications can be classified as \textit{networked control systems (NCSs)}, i.e., feedback control loops that are closed over a communication network. In contrast to conventional control theory, NCSs accommodate at least one link in the feedback loop that is not ideal and therefore has a direct effect on the underlying application's performance. In particular, as the scarcity of network resources increases, e.g., due to a large number of users in the network, those components located at the other end of the imperfect communication link may retain outdated information due to delays and packet loss. As a result, the control performance degradation is inevitable and it may lead to the destabilization of the system and possible physical damage in the environment or injuries of human operators.

	One of the main differences of NCSs from traditional communication systems is that their information exchange takes place to complete a certain task in physical environment. To give an example, the communication between the sensors of an autonomous car and its electronic control unit aims at preventing potential accidents or improving driving experience. In a network comprised of such applications, the performance of the overall system is not measured by the amount of bits correctly transmitted over a noisy channel, but rather by the efficiency in completion of corresponding tasks \cite{mostaani2021taskoriented}. Hence, while the communication networks go through a paradigm shift revisiting the common Shannon theory, the wireless system design can not be considered as task-agnostic. Indeed, identifying the most relevant information in order to achieve the defined control goals should be of great importance when making decisions in the communication network. Especially, as the scarcity of such resources as bandwidth, energy, increases with the growing number of connected machines, it becomes crucial to focus on ``what" to transmit instead of ``how" to transmit.  In fact, it is foreseen that already the 6G cellular networks should address this challenge by including significance and effectiveness aspects of information into network design \cite{calvanese20216g}.
	
	In order to identify the significance of information, one has to study various properties of information. One of these properties is freshness. In order to quantify information freshness at a destination, the metric \textit{age of information (AoI)} has been proposed \cite{kaul2012real}. AoI is defined as the elapsed time since the generation of the most recent received packet, thus it captures the staleness of  information at the destination. It is an application layer metric measured from the perspective of the receiver, where it is utilized in order to fulfill a certain task, such as real-time monitoring and control. Its simplicity and the ability to combine end-to-end delay and packet loss makes it a very suitable metric for time-sensitive applications and cross-layer network design.
	There are extensive theoretical research papers on minimizing AoI in a variety of contexts. Works \cite{kadota2016minimizing, kadota2018scheduling, hsu2018scheduling} focus on finding age-optimal scheduling policies in single-hop networks, whereas \cite{maatouk2018age, vikhrova2020age, talak2017minimizing} consider multi-hop topology. Moreover, \cite{kaul2012status, costa2014age, kam2016controlling} derive AoI performance when different queuing disciplines are applied. They show that replacing outdated packets with newly generated ones in the transmission queue is beneficial w.r.t. information freshness.
	
	AoI is successful at capturing the significance of data up to a certain degree by measuring their timeliness property. However, in task-oriented communication systems such as connected robotics, the received status update packets are further processed for estimation, control and other context dependent tasks. Consequently, the overall performance relative to system's design purpose is not only affected by messages' timeliness but also by their significance and content. Therefore, \textit{semantics of information}, which is defined as the significance of data relative to their transmission purpose in \cite{uysal2021semantic}, is envisioned to play a key role for future networked systems and especially in task-oriented communication systems.
	\subsection{Related Work}
	\subsubsection{Theoretical Research}Adoption of properties of information as in the case of AoI has given rise to the adoption of metrics beyond AoI such as non-linear age, ``value of information" (VoI) and ``age of incorrect information" \cite{kosta2017age, maatouk2020age, kosta2021age} in single user scenarios. 
	The notion of non-linear aging has been refined for control applications by using system dependent parameters \cite{ayan2019aoi}. In their work, authors define the VoI as a function of control system parameters and employ it for the wireless resource allocation problem. They show that it outperforms the AoI when it comes to control performance, although the network-wide information freshness is decreased. Moreover, \cite{ayan2021age} focuses on improving control performance for queuing systems in which an optimal distribution of the total available service rate is found. 
	\cite{champati2019performance} is another example to the adoption of AoI-based functions in NCSs domain. It suggests an optimal sampling policy for a single-user scenario that is minimizing AoI-based functions. However, \cite{ayan2019aoi, champati2019performance, ayan2021age} assume either zero end-to-end delay or base their evaluation on constant service rate. Additionally, \cite{walsh2001scheduling} studies the centralized scheduling problem of NCSs, the authors of which propose a greedy scheduling protocol based on the estimation error caused by the communication network. However, they assume global knowledge of the instantaneous error at the scheduler which is not a feasible assumption when it comes to the practical implementation of the proposed algorithm. All of the aforementioned works provide valuable insights into AoI and NCSs domains. However, they do not capture important system related complications of practical deployment.

	\subsubsection{Systems Research}
	The vast majority of previous work on AoI and NCSs, including but not limited to the papers mentioned in the previous subsection, has been theoretical.  One of the main reasons for this has been the significant effort demanded for the modification of the communication stack, which has hindered the validation of proposed solutions on hardware. 
	
	\cite{sonmez2018age, barakat2019how, beytur2019measuring} are the first examples of practical AoI research, which measure AoI performance in real-life connections without any modifications in the communication stack. \cite{shreedhar2019age} is the first work known to us towards AoI-aware customization in which the authors propose a transport layer protocol for increased information freshness. However, their implementation is limited to rate control and lower layers are transparent to the source, thus are left unmodified.
	
	The recent increasing popularity of softwarization in networking, especially the introduction of software-defined radio (SDR) platforms in wireless research, has lowered the barrier to go beyond transport layer customization on real hardware. \cite{kadota2020age, han2020software, ayan2021experimental} propose customized solutions using SDRs for improved network-wide information freshness. To the best of our knowledge, these are the only AoI research papers that propose a customized MAC layer solution using real-world equipment. In \cite{han2020software}, authors propose an AoI-threshold based random access protocol for wireless networks that reduces the mean AoI when compared to the well-known slotted ALOHA protocol. \cite{kadota2020age} proposes a contention-free wireless MAC protocol implemented on SDR platforms and shows that their framework outperforms a standard WiFi network w.r.t. information freshness. \cite{ayan2021experimental} considers multiple NCSs sharing a wireless communication network. It compares the network-wide AoI and control performance when different queuing disciplines such as last come first serve (LCFS) and first come first serve (FCFS) are employed. The authors conclude that in their considered scenario, LCFS discipline performs significantly better as the resource scarcity of the network increases. However, they assume homogeneous type of control systems and their work is limited to a performance comparison between different queuing disciplines. In contrast to our work, it does not suggest any customized task-oriented MAC protocol towards an improvement in the AoI or control performance.
	
	\subsection{Main Contributions}
	We believe that systems research combining theory and practice may reveal some unrealistic or strict assumptions that are made in theoretical research papers. In addition, it points out possible improvements in simulations and system modeling towards a design closer to the reality. Moreover, as we are going to present in the results section (e.g., Fig. \ref{fig:ra_mean_aoi}), it reveals possible sources of mismatch between analytical and experimental results.
	
	Our main goal is to contribute to the practical AoI and NCSs research with an extensive experimental study. In a network comprised of multiple heterogeneous control sub-systems, we realize selected wireless MAC protocols from the literature that have been designed to increase information freshness and control performance. We compare the performance of these protocols for different key performance indicators (KPIs) and detect their strengths and weaknesses in a practical setting for increasing number of loops. In particular, we measure KPIs such as mean AoI, mean squared estimation error and other cost metrics from the NCSs domain. Our results and findings are based on real-world measurements consisting of up to $15$ SDRs programmed in C++ using GNU Radio software radio framework \cite{gnuradio}.
	
	In addition to the realization of existing solutions, we propose a new protocol that combines the core ideas and strengths of two prior works, namely, \cite{ayan2019aoi} and \cite{kadota2020age}. In particular, it is a polling-based, \textit{control-aware} MAC protocol, a protocol that takes control system parameters into account, and uses it for prioritization of sub-systems. We show that our proposed solution is able to outperform its closest competitor by up to 21\% with respect to control performance. To the best of our knowledge, this is the \textit{first practical work} implementing a control-aware MAC protocol on real hardware.
	
	As a second main improvement over the state-of-the-art (SotA), we propose a normalization technique that can be used together with SotA solutions utilizing control related metrics such as in our previous work \cite{ayan2019aoi}. The suggested enhancement facilitates the applicability of system dependent metrics particularly in scenarios with heterogeneous control systems. Additionally, in a case study involving multiple control loops of inverted pendulum type, we demonstrate why the existing metric from \cite{ayan2019aoi} is not capable of capturing the urgency of transmission in its original form. Furthermore, we show that when the new metric is employed in combination with our proposed polling-based scheduling algorithm, the inverted pendulums are successfully stabilized with $15$ feedback control loops in the network.

	The remainder of this paper is outlined as follows. In section \ref{sec:scenario}, we introduce the considered control model and network scenario. In section \ref{sec:aoiandcontrol}, we discuss how control performance depends on AoI and formulate the problem statement for wireless resource scheduling. Section \ref{sec:protocols} gives an overview of the implemented MAC protocols from the SotA and introduces our proposed solution. The details of our implementation are given in section \ref{sec:implementation}. In section \ref{sec:results} we provide and discuss our experimental results. Section \ref{sec:conclusion} concludes the paper.
	
	\subsection{Notation}
	$\mathbb{N}_0$ denotes the natural numbers including zero. The positive natural numbers are denoted by $\mathbb{N}^+$. Throughout the paper, matrices are denoted by capital letters in bold font, i.e., $\bm{M}$, whereas small letters are used for vectors, i.e., $\bm{v}$. Transpose of a matrix $\bm{M}$ is given as $\bm{M}^T$. Moreover, $\bm{M}^p$ is the $p$-th power of a matrix $\bm{M}$.
	
	\section{Scenario and Background on Remote Control}
	\label{sec:scenario}
	\subsection{Network}
	\begin{figure}[t]
		\centering
		\scalebox{0.9}{
			\input{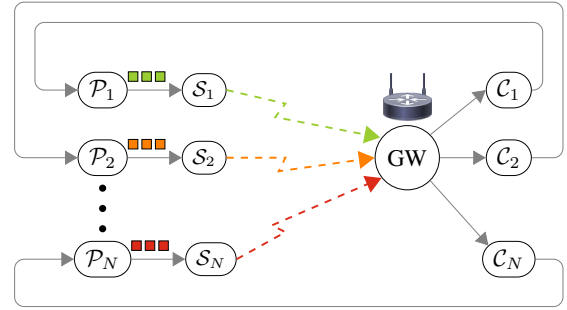}
		}
		\caption{The considered scenario with $N$ feedback control loops closed over a shared wireless link. Each SDR $\sdr_i$ is responsible for transmitting status update packets of plant $\plant_i$ to the gateway (GW) node from where it is forwarded to the corresponding controller $\controller_i$. Solid arrows represent ideal links between components of a feedback loop.}
		\label{fig:scenario}
	\end{figure}
	We consider $N$ heterogeneous feedback control loops closed over a shared wireless channel. Each loop consists of a plant and a controller. The plant $\mathcal{P}_i$ of the $i$-th sub-system is the entity that is to be controlled by the controller $\mathcal{C}_i$ where the goal of $\mathcal{C}_i$ is to drive $\mathcal{P}_i$ to a desired state. 
	
	The controller is able to observe the plant state via the shared wireless channel where each packet containing a single status update is transmitted by a SDR $\sdr_i$ to a gateway (GW) from where it is forwarded to $\controller_i$. 
	In this work, we consider the controller-to-plant link to be ideal. The camera-based control of an inverted pendulum can be named as a practical example to such a topology\footnote{An inverted pendulum is explained in Sec. \ref{subsec:casestudy} in detail. Fig. \ref{fig:invertedpendulum} depicts an inverted pendulum.}. While the camera observes the system remotely and transmits real-time state measurements over a wireless network, the plant and the controller are co-located. As a result, the topology can be viewed as $N$ source nodes contending for channel resources to transmit their status updates over the shared single-hop wireless communication link. Fig. \ref{fig:scenario} depicts the resulting network topology that is considered throughout the following sections. 
	
	From the theoretical AoI research \cite{kaul2012status}, we know that under the  assumption that status is Markovian, having  received an update, the controller does not benefit from receiving an older observation. Thus, older packets are considered as obsolete  and  non-informative. Consequently, in our framework, as well as in the following analysis, we assume that each $\sdr_i$ discards any older packet upon the generation of a new state information. This queuing discipline is referred to as LCFS in the literature and has been proven to outperform practical implementations employing FCFS \cite{kadota2020age, ayan2021experimental}.
	
	\subsection{Control}
	In our setup, we employ digital representation of control sub-systems that are running as independent processes parallel to their communication counterparts, i.e., SDRs. The behavior of the control loops are modeled as discrete time linear time-invariant (LTI) systems. That is, the system state of $\mathcal{P}_i$ varies over time in discrete steps with a constant period of $T_{i, s}$. In other words, two consecutive discrete time steps $t$ and $t+1$ are $T_{i,s}$ seconds apart in continuous time. 
	
	The system state of each sub-system $i$ follows the state-space representation:
	\begin{equation}
	\statevec_i[t+1] = \statemat_i \statevec_i[t] + \inputmat_i \inputvec_i[t] + \noisevec_i[t].
	\label{eq:statespace}
	\end{equation}
	Here, $\statevec_i \in \mathbb{R}^{n_i}$ and $\inputvec_i \in \mathbb{R}^{m_i}$ are column vectors denoting the plant state and control input, respectively. $\statemat_i \in \mathbb{R}^{n_i \times n_i}$ is the time-invariant system matrix, which defines the relationship between the current system state $\statevec_i[t]$ and the next state $\statevec_i [t+1]$. Moreover, $\inputmat_i \in \mathbb{R}^{n_i \times m_i}$ is the time-invariant input matrix, which defines the effect of the control input on the next system state. The noise vector $\noisevec_i \in \mathbb{R}^{n_i}$ is considered to be independent and identically distributed (i.i.d.) according to a zero-mean Gaussian distribution with diagonal covariance matrix $\bm{\Sigma}_i \in \mathbb{R}^{n_i \times n_i}$, i.e., $\bm{w}_i \sim \mathcal{N}\left( \bm{0}, \bm{\Sigma}_i \right)$.
	
	It is important to emphasize that every $T_{i, s}$ seconds the system state $\statevec_i[t]$ is updated according to \eqref{eq:statespace} from $\statevec_i[t]$ to $\statevec_i[t+1]$. Between the two consecutive update instances, which are also referred to as \textit{sampling} instances in the literature, the state is considered to be constant. However, we assume that $T_{i, s}$, i.e., the \textit{sampling period} is selected small enough such that \eqref{eq:statespace} sufficiently approximates the continuous time behavior of the real control system. This is a well-established approach in control theory textbooks to model continuous time systems in discrete time \cite{astrom2008feedback}. In order to simplify the following analysis and the protocol design, we have selected the sampling period to be equal among all control sub-systems, i.e., $T_{i, s} = T_s, \, \forall i$.
	
	The input signal $\inputvec_i[t]$ is calculated based on the observation history available at $\controller_i$. However, as the state information is delivered via the shared wireless $\sdr_i$-to-$\controller_i$ link, packet collisions and the wireless resource scarcity lead to the fact that only a subset of the generated packets at each source is successfully delivered to $\controller_i$. In addition, those are delivered with a certain non-negligible end-to-end delay caused by the communication stack between the plant and controller processes. As a result, the information at the controller is outdated and the network-induced information staleness at $\controller_i$ leads to inaccurate control inputs, hence to the degradation of the overall control performance. 
	
	To reduce the negative effects of information staleness, each controller $\controller_i$ employs an estimator to estimate the current state $\statevec_i[t]$ remotely. In particular, given the freshest information $\statevec_i[\nu_i(t)]$ generated at sampling period $\nu_i(t)$ and received by $\controller_i$ until the beginning of the $t$-th sampling period, the controller of sub-system $i$ estimates the current state based on its expected value as:
	\begin{align}
	\bm{\hat{x}}_i[t] &\triangleq \mathbb{E}\left[\statevec_i[t] ~|~ \statevec_i[\nu_i(t)] \right] \nonumber \\
	&= \statemat_i^{\Delta_i[t]} \statevec_i[\nu_i(t)] + \sum_{q =1}^{\Delta_i[t]} \statemat_i^{q - 1} \inputmat_i \inputvec_i[t - q],
	\label{eq:estimator}
	\end{align}
	with $\Delta_i[t] \triangleq t - \nu_i(t), \quad  t, \nu_i(t) \in \mathbb{N}_0, ~\forall i$. The model of the remote state estimation is taken from \cite{ayan2021age}, in which the authors provide the proof of \eqref{eq:estimator}. Similar to their work, $\Delta_i[t]$ is defined as the number of elapsed sampling periods since the generation of the freshest state information available at $\controller_i$, thus \textit{age of information (AoI)} in the unit of $T_s$. The age model is discussed in detail later in Sec. \ref{subsec:aoi}.
	
	We assume that the design of the optimal controller is done independently of the network and prior to deployment of control loops. Therefore, we select the commonly used \textit{linear quadratic regulator (LQR)} for controller design that aims to minimize the infinite horizon quadratic cost function:
	\begin{equation}
	\label{eq:lqg_cost}
	F_i \triangleq \limsup_{T \rightarrow \infty} \mathbb{E}\left[ \dfrac{1}{T} \sum_{t=0}^{T - 1} (\boldsymbol{x}_i[t])^T \boldsymbol{Q}_i \boldsymbol{x}_i[t] +  (\boldsymbol{u}_i[t])^T \boldsymbol{R}_i \boldsymbol{u}_i[t] \right].
	\end{equation}
	The matrices $\bm{Q}_i$ and $\bm{R}_i$ are symmetric positive semi-definite weighting matrices of appropriate dimensions that are used to penalize the state error and control effort, respectively. Throughout the paper we assume the set-point to be zero, therefore the state $\bm{x}_i$ is essentially its deviation from the desired value. In control theory textbooks, $F_i$ is referred to as \textit{linear-quadratic-Gaussian (LQG)} cost function.
	
	The controller obtains the control input by following a linear, time-invariant \textit{control law} \cite{astrom2008feedback}:
	\begin{equation}
	\label{eq:controllaw}
	\inputvec_i[t] = - \bm{L^*}_i \bm{\hat{x}}_i[t],
	\end{equation}
	where $\bm{L^*}_i \in \mathbb{R}^{m_i \times n_i}$ is the optimal state feedback gain matrix. The calculation of $\bm{L^*}_i$ follows by solving the discrete time algebraic Riccati equation:
	\begin{equation}
	\label{eq:riccati}
	\boldsymbol{P}_i = \boldsymbol{Q}_i + \boldsymbol{A}_i^T \big( \boldsymbol{P}_i - \boldsymbol{P}_i \boldsymbol{B}_i \big( \boldsymbol{R}_i + \boldsymbol{B}_i^T \boldsymbol{P}_i \boldsymbol{B}_i \big)^{-1} \boldsymbol{B}_i^T \boldsymbol{P}_i \big) \boldsymbol{A}_i,
	\end{equation}
	with the solution:
	\begin{equation}
	\label{eq:riccatisolution}
	\boldsymbol{L^*}_i = \big( \boldsymbol{R}_i + \boldsymbol{B}_i^T \boldsymbol{P}_i \boldsymbol{B}_i \big)^{-1} \boldsymbol{B_i}^T\boldsymbol{P}_i \boldsymbol{A}_i.
	\end{equation}
	
	In simple words, the operation of the controller can be summarized as follows: After each estimation step performed according to \eqref{eq:estimator}, the controller uses $\bm{\hat{x}}_i[t]$ to determine the control input following the control law in \eqref{eq:controllaw}. The resulting $\inputvec_i[t]$ is then applied to $\plant_i$ during the next sampling period $t$. The freshest packet that has been received until the end of the $t$-th sampling period is then used for the estimation of $\statevec_i[t+1]$ and the next control input is obtained analogously.
	
	We would like to mention that $\boldsymbol{L^*}_i$ is the optimal matrix minimizing the LQG cost $F_i$ without the consideration of the network. However, as the authors of \cite{maity2022optimal} show in Corollary ~1, the controller with the conditional state estimation as in \eqref{eq:estimator} and the optimal feedback matrix $\boldsymbol{L^*}_i$ obtained by solving the standard LQG problem, in fact leads to the optimal control law as in \eqref{eq:controllaw} if the network is prone to delays and dropouts. The effects of the network imperfections are reflected in the estimation process.
	
	\section{Information Staleness and Effects on Control Performance}
	\label{sec:aoiandcontrol}
	\subsection{Age of Information}
	\label{subsec:aoi}
	As described in Sec. \ref{sec:scenario}, each controller obtains a remote state estimate based on the freshest information available. However, especially in a real network as in our considered scenario, it is common to observe network-induced delays originating from processing and transmission. In addition, part of the generated data is either discarded in the transmission queue or ``lost" in the channel due to bad link quality or simultaneous access. All these effects combined lead to information staleness and in consequence to inaccurate state estimation. In that case, the controller's actions become sub-optimal which lead to increased state deviation from the equilibrium. This makes the controller put more effort into driving the state back to the desired value. As a result, the control cost $F_i$ given in \eqref{eq:lqg_cost}, which is characterized jointly by the state error and the control effort, increases.
	
	As in \eqref{eq:estimator}, let $\statevec_i[\nu_i(t)]$ be the most recent information available at $\controller_i$ that denotes the system state at $\nu_i(t)$, where $\nu_i(t) < t$ holds\footnote{$\nu_i(t)$ is always smaller than $t$ because in our implementation, the calculation of $u_i[t]$ happens directly subsequent to sampling. As it is infeasible to have ``almost zero" delay in a practical setup, in our mathematical model, we do not allow the equality case, i.e., $\nu_i(t) < t, ~ \forall t$.}. From Sec. \ref{sec:scenario} we know that the state of our plant process only changes with discrete and constant intervals over time. Therefore, since our goal is to quantify the age of the freshest information, we are interested in the difference between the current time step $t$ and the generation time step $\nu_i(t)$, i.e., AoI, in units of $T_s$.
	\begin{figure}[t]
		\centering
		\scalebox{0.8}{
			\begin{tikzpicture}
			\begin{axis}[
			axis lines = left,
			xlabel={Discrete time step $t$},
			ylabel={Age of information $\Delta_i[t]$},
			xmin=0, xmax=22,
			ymin=0, ymax=11,
			grid=both,
			minor x tick num=3,
			minor y tick num=1,
			xtick pos=bottom,
			ytick pos=left,
			xtick={1,5,9,13,17,21},
			ytick={0,2,4,6,8,10,12},
			legend pos=north west,
			ymajorgrids=true,
			grid style=dotted,
			]
			
			\addplot[
			color=blue,
			mark=*,
			mark options={solid},
			dashed
			]
			coordinates {
				(1,1)(2,2)(3,3)(4,4)(5,2)(6,1)(7,2)(8,3)(9,4)(10,5)(11,6)(12,7)(13,8)(14,9)(15,1)(16,2)(17,3)(18,4)(19,5)(20,1)(21,2)
			};
			\node[anchor=north west, fill=none] (source) at (axis cs:-0,10.5){\color{red}{\ Reception of $\statevec_i[3]$ and $\statevec_i[5]$ by $\controller_i$}};
			\node (destination1) at (axis cs:5, 2.5){};
			\node (destination2) at (axis cs:6, 1.5){};
			\draw[-triangle 45, thick, red](source)--(destination1);
			\draw[-triangle 45, thick, red](source)--(destination2);
			\end{axis}
			\end{tikzpicture}
		}
		\caption{Example evolution of discrete time AoI of a sub-system $i$, i.e., $\Delta_i[t]$ recorded during a real-world measurement using our experimental platform. We observe four reduction of AoI, namely at $t=\{5, 6, 15, 20\}$. Note that the dashed line does not represent the evolution of AoI in continuous time.} 
		\label{fig:AoI}
	\end{figure}
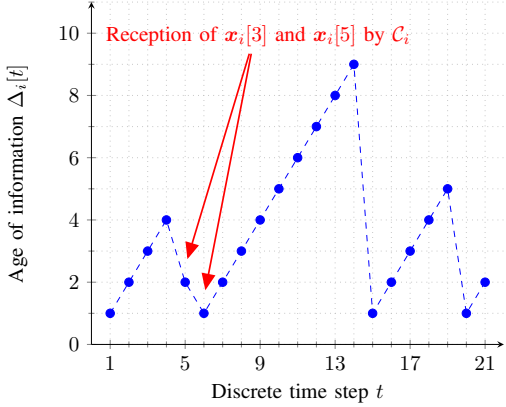
	
	For better understanding, we present the evolution of AoI during a $20$ sampling periods long measurement. Fig. \ref{fig:AoI} shows the AoI at one of the controllers monitoring the state of its respective plant via a shared channel. The figure plots $\Delta_i[t]$ over $t=\{1, 2, \dots, 21\}$ where the initial AoI is one, i.e., $\Delta_i[1] = 1$ or $\nu_i(1) = 0$. We observe that AoI drops four times during our measurement, at $t = \{5, 6, 15, 20\}$ due to a successful update. First we see a linear increase up to $\Delta_i[4] = 4$ with a slope of $1$ indicating no new reception prior to the beginning of the fourth sampling period, i.e., $\nu_i(t) = 0$ for $t \leq 4$. Until $\controller_i$ begins with the calculation of $\inputvec_i[5]$, one or more packets have been successfully decoded by the controller, with the freshest packet containing the system state $\statevec_i[3]$. In other words, two sampling periods have elapsed until the new update has successfully been used by $\controller_i$ when $\inputvec_i[5]$ is obtained and AoI drops to $2$. During the next sampling period, $\controller_i$ receives $\statevec_i[5]$ which leads to $\nu_i(6) = 5$. Similarly, it follows that $\nu_i(t) = 5$ for $6 \leq t \leq 14$ and $\nu_i(t) = 14$ for $15 \leq t \leq 19$. It is important to emphasize that the dashed line connecting the round markers at discrete time steps of $t$ does not represent the continuous time behavior of $\Delta_i[t]$ as it is only defined at sampling instances.

	\subsection{Estimation Error}
	\label{subsec:estimationerror}
	Let us consider two real-time processes that are sampled with the same sampling frequency of $T_s = 10$ milliseconds, e.g., the temperature of an office room which may not vary a lot over long time periods and the location of an unmanned aerial vehicle (UAV) that is highly mobile. Moreover, suppose that we are monitoring the states of these two plants via a communication network and we are only able to transmit the latest system state only once in every 1000 packets, i.e., once per 10 seconds.
	
	If we are dealing with different classes of applications as in this toy example, we can intuitively see that the AoI is incapable of capturing the uncertainty growing at the monitor over time between two consecutive status updates. In other words, the value of transmitting the next packet when the AoI reaches $1000$ is different for the considered applications as they are unlike in state dynamics.
	
	One way of capturing the uncertainty at the destination monitoring heterogeneous sources is to use the estimation error. The \textit{estimation error} is defined as the difference between the real system state and the estimated system state, i.e.:
	\begin{align}
	\bm{e}_i[t] &\triangleq \statevec_i[t] - \bm{\hat{x}}_i[t] \nonumber \\
	&= \sum_{d = 1}^{\Delta_i[t]} \bm{A}_i^{d - 1} \bm{w}_i[t - d].
	\label{eq:estimationerror}
	\end{align}
	The closed form equation for $\bm{e}_i[t]$ can be obtained by subtracting \eqref{eq:estimator} from \eqref{eq:statespace}.
	The mean squared error (MSE), which can be derived from the estimation error, is widely used in the literature to quantify estimation performance. It can be obtained by taking the expectation of a quadratic form as:
	\begin{equation}
	MSE_i[t] \triangleq \mathbb{E}\left[(\bm{e}_i[t])^T \bm{e}_i[t]\right].
	\end{equation}
	In \cite{ayan2019aoi}, we derive the MSE as a function of AoI for the same model of an NCS as in this work:
	\begin{equation}
	MSE_i[t] = \sum_{d=1}^{\Delta_i[t]} \mathsf{tr}\left( (\statemat_i^T)^{d - 1} \statemat_i^{d-1} \bm{\Sigma}_i \right),
	\label{eq:MSEage}
	\end{equation}
	with the trace operator $\mathsf{tr}(.)$. Here, $\statemat_i$, $\bm{\Sigma}_i$ are defined as in \eqref{eq:statespace}. \eqref{eq:MSEage} maps the instantaneous AoI $\Delta_i[t]$ to MSE that strongly depends on control system parameters such as the system matrix and noise covariance matrix. Note that these parameters are time-invariant and $\Delta_i[t]$ is the only time-dependent variable in the equation.

	$\bm{e}_i[t] \in \mathbb{R}^{n_i}$ is a multi-variate random variable (RV) defined as the deviation of the system state from its expectation. The first property of $\bm{e}_i[t]$ is that it is a zero-mean multivariate RV, i.e., $\mathbb{E}\left[ \bm{e}_i[t] \right] = \bm{0}$, with $\bm{0}$ being a column vector of   length $n_i$ that contains only zeros. This can easily be shown by taking the expectation of the right hand side (RHS) of \eqref{eq:estimationerror} and applying the linearity property of expectation. Moreover, $\bm{e}_i[t]$ is a normally distributed multi-variate RV since each addend in \eqref{eq:estimationerror} is a linear transformation of the multivariate normal RV $\bm{w}_i[t - d] \sim \mathcal{N}\left(\bm{0}, \bm{\Sigma}_i \right)$ with $1 \leq d \leq \Delta_i[t]$. In fact, each addend follows a normal distribution with the covariance matrix $\bm{\Sigma}_{d} = \bm{A}_i^{d - 1} \bm{\Sigma}_i (\bm{A}_i^{d - 1})^T$.
	
	\begin{proof}
		Given any $d \geq 1$, the $d$-th addend of \eqref{eq:estimationerror} is $\bm{y}_d[t] = \bm{A}_i^{d - 1} \bm{w}_i[t - d]$ with $\bm{y}_d[t] \in \mathbb{R}^{n_i}$ and $\mathbb{E}\left[ \bm{y}_d[t]\right] = \bm{0}$. The covariance $\bm{\Sigma}_{d}$ can be written as:
		\begin{align*}
			\bm{\Sigma}_{d} &\triangleq \mathbb{E}\left[ (\bm{y}_d - \mathbb{E}\left[ \bm{y}_d\right]) (\bm{y}_d - \mathbb{E}\left[ \bm{y}_d\right])^T \right] \\
			&= \mathbb{E}\left[ \bm{A}_i^{d-1} \bm{w}_i[t - d] (\bm{w}_i[t - d])^T (\bm{A}_i^{d-1})^T  \right] \\
			&=  \bm{A}_i^{d-1} \mathbb{E}\left[\bm{w}_i[t - d] (\bm{w}_i[t - d])^T  \right] (\bm{A}_i^{d-1})^T  \\
			&= \bm{A}_i^{d-1} \bm{\Sigma}_i (\bm{A}_i^{d-1})^T
		\end{align*} 
	\end{proof}
	The overall estimation error $\bm{e}_i[t]$, which is comprised of $d$ independent addends, i.e., $\left \{\bm{y}_d[t] : 1 \leq d \leq \Delta_i[t] \right \}$, is characterized by the multivariate normal distribution
	$\bm{e}_i[t] \sim \mathcal{N}\left(\bm{0}, \bm{\Sigma}_e \right)$. Since we are able to sum up the covariance matrices as the individual addends are independent RVs, it holds that:
	
	\begin{equation}
		\bm{\Sigma}_e = \sum_{d = 1}^{\Delta_i[t]} \bm{A}_i^{d-1} \bm{\Sigma}_i (\bm{A}_i^{d-1})^T.
		\label{eq:covariance_error}
	\end{equation}
	Here it is important to emphasize that an increase in $\Delta_i$ leads to a new positive semi-definite addend on the RHS. Note that if $\bm{A}_i$ is a scalar, this would correspond to an increase in the variance of the distribution that $\bm{e}_i$ follows. Let us illustrate this with a numerical example that considers a scalar loop with $\bm{A}_i = 1.2$ and $\bm{\Sigma}_i = 1.0$.
	

	\begin{figure}[t]
		\centering
		\includegraphics[width=0.5\textwidth]{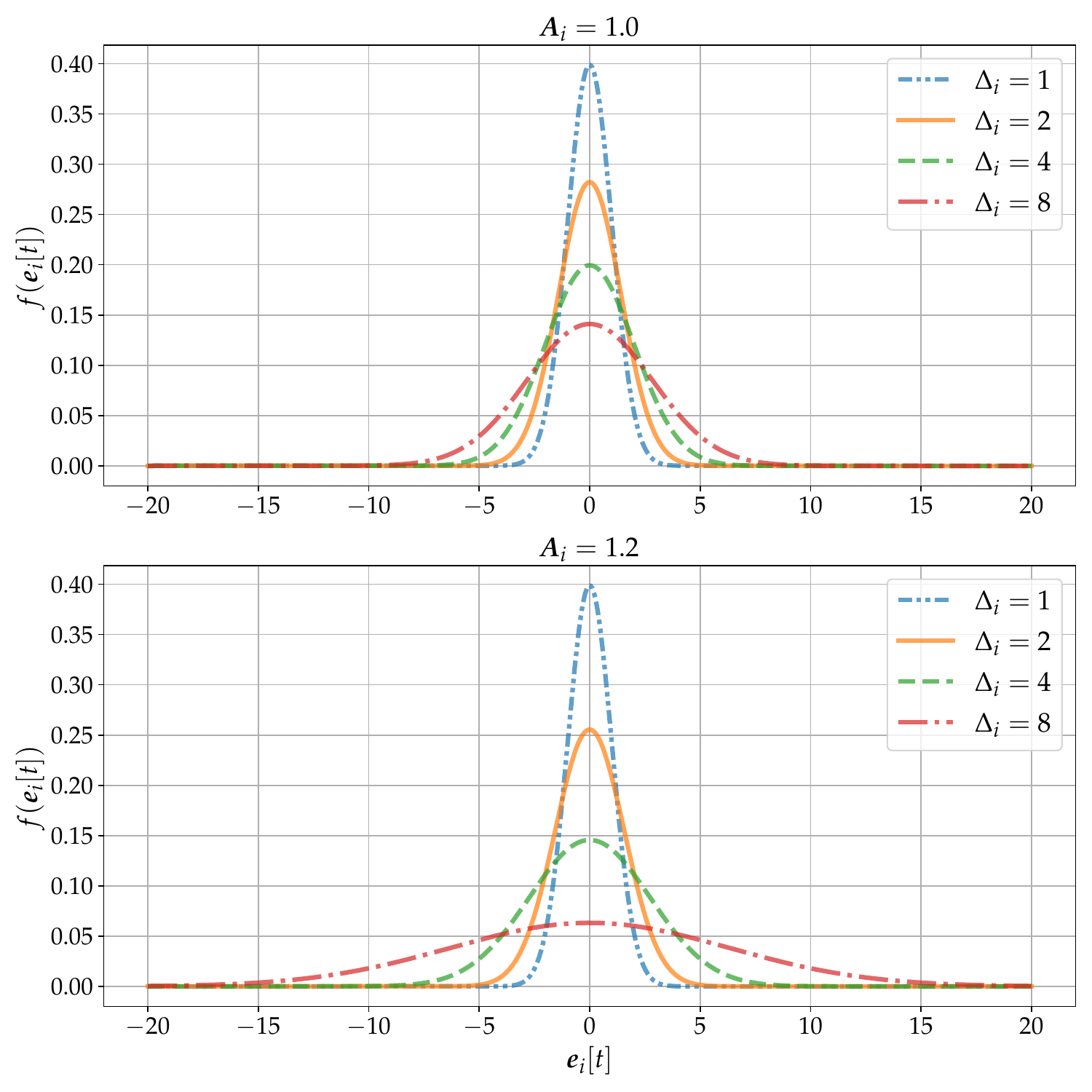}
		\caption{The probability density function of the estimation error $f(\bm{e}_i[t])$ for varying AoI values. The distribution is characterized by $\bm{e}_i[t] \sim \mathcal{N}\left(\bm{0}, \bm{\Sigma}_e \right)$  with $\bm{\Sigma}_e = \sum_{d = 1}^{\Delta_i[t]} \bm{A}_i^{d-1} \bm{\Sigma}_i (\bm{A}_i^{d-1})^T$. Here, $\bm{A}_i = \{1.0, 1.2\}$ and $\bm{\Sigma}_i = 1.0$ are used.}
		\label{fig:error_distr}
		
	\end{figure}
	 Fig. \ref{fig:error_distr} depicts the probability density function (PDF) of estimation error for different control systems when the AoI ranges from $1$ to $8$.
	 From the figure, one can see how the PDFs become more stretched as information staleness at the estimator increases. Put differently, if we consider the estimation error as the deviation of the estimated state from the actual system state, the uncertainty of our estimation about the remote state grows with the increasing $\Delta_i$. It is important to mention that this uncertainty does not grow at the same speed for every control application as information gets outdated. In fact, the sub-system with $\bm{A}_i = 1.2$ depicted at the bottom has a much wider distribution of the squared error at $\Delta_i= 8$ than the one with $\bm{A}_i = 1.0$ shown at the top. The figure can be interpreted as an illustration of how the significance of transmitting the next status update relates to the freshness property of information and to its context, i.e., who is sending and receiving the information, what is the purpose of conveying this information, etc. In our toy example illustrated in the figure, the context of communication is defined by the goal of uncertainty reduction at two destinations that are monitoring two remote processes with distinct system dynamics.
 	
 	\begin{figure*}[t]
 		\centering
 		\includegraphics[width=.8\textwidth]{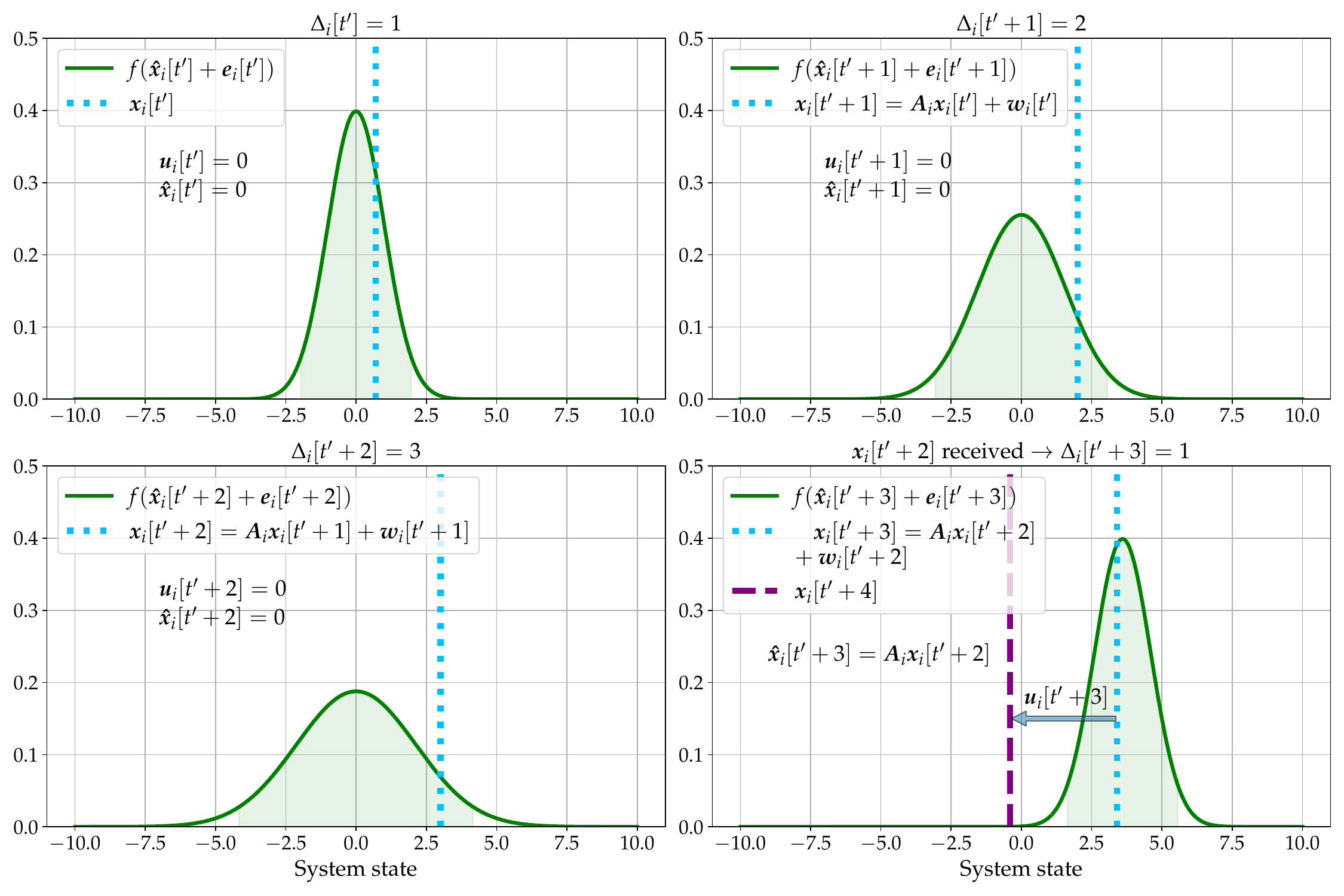}
 		\caption{An example snapshot of the system state $\bm{x}_i[t]$, control input $\bm{u}_i[t]$ and estimated state $\bm{\hat{x}}_i[t]$. The figures illustrate how the state drifts away from the reference value due to missing status updates about recent changes. Please notice that the distribution of the estimation error is more stretched as $\Delta_i$ increases.}
 		\label{fig:state_evol}
 		
 	\end{figure*}

 	Although the estimation error is not a direct measure of control performance, it has a strong effect on the accuracy of control inputs. That is, with the growing uncertainty at each controller $\controller_i$, the applied control inputs become sub-optimal due to the deviation between $\bm{x}_i[t]$ and $\bm{\hat{x}}_i[t]$. Consequently, $\bm{u}_i[t]$ is not able to drive the state towards the reference value correctly. This causes an increase in the overall control cost $F_i$ since the state grows as well as the control effort. This phenomenon is shown in Fig. \ref{fig:state_evol}, where the relationship between a wrong state estimate and an imperfect control input is illustrated.
 	
 	In Fig. \ref{fig:state_evol}, we see an example interplay between the AoI, system state, its estimation and the control input. The controller has an outdated information and expects $\bm{x}_i[t']$ to be correctly driven to the equilibrium point of $\bm{x}_i = 0$. As the AoI increases further, the controller does not take any immediate action due to the lack of recent information, i.e., $\bm{u}_i[t] = 0$ for $t \in [t', t' + 2]$. Only upon the reception of a new update, the controller improves its estimation and applies a non-zero control input at $t = t' + 3$ to drive the state back to zero. Both the state deviation and the following control effort contribute to the LQG cost and lead to a degradation in the control performance.

	\subsection{Task-Oriented Communications and Problem Statement}
	The calculation of the optimal state feedback gain matrix $\bm{L^*}_i$ from \eqref{eq:controllaw} is done by assuming ideal communication links between the components of a feedback control loop. However, this contradicts with our considered scenario, in which the state observations are sent over a physical wireless link. Therefore, to limit the deviation of controller design from optimality w.r.t. LQG cost, the network should aim at reducing the error between the actual and estimated states that is induced by the imperfect communication links within sensor-controller pairs.
	
	As the network consists of multiple control sub-systems and the  available bandwidth is limited, we need to identify the most relevant transmissions and fit them into the available network resources to improve performance. Considering the fact that in our setup the source-destination pairs are represented by sensors and controllers, that would correspond to selecting the highest instantaneous uncertainty reduction at controller in case of a successful transmission, hence scheduling the user with the highest MSE\footnote{Note that the uncertainty reduction happens only if the transmission is successful. This requires the consideration of packet success probability. In Sec. \ref{subsec:contentionfreeprotocols} we discuss in detail how the link reliability is incorporated into scheduling decisions in our setup.}.
	
	On the contrary, the control theory uses LQG cost $F_i$ as a metric to quantify the success level in accomplishing the control goal. Although it is challenging to analytically formulate the exact relationship between MSE and LQG cost, they are strongly intertwined as discussed in previous section. Having that said, our approach exploits the indirect relationship between the estimation and control performances. In other words, by reducing the overall MSE in the network, we expect to reduce the LQG cost and thus improve the quality of control\footnote{This expectation is based on the results of a previous work \cite{ayan2021age} that studies a FCFS discrete time queue in a simulation-based setup.}. Therefore, our final goal is to implement a customized wireless medium access control (MAC) protocol $\bm{\pi}$ on SDRs such that the average LQG cost per control sub-system is minimized, i.e.:
	\begin{equation}
	\bm{\pi} = \arg \min_{\bm{\pi}} ~ \dfrac{1}{N}\sum_{i=1}^{N}F_i^{\bm{\pi}},
	\end{equation}
	with $F_i^{\bm{\pi}}$ being the linear quadratic cost when $\bm{\pi}$ is employed. Section \ref{sec:protocols} presents two examples to such wireless MAC protocols using MSE in the context of control-oriented communications. While the first is an existing protocol from the literature, the second is a new solution firstly proposed in this work.

	
	\section{MAC Protocols for Real-time NCSs}
	\label{sec:protocols}
	In this section, we introduce various, selected MAC protocols that we have implemented and tested in our experimental framework. First, we explain three existing contention-based protocols in \ref{subsec:randomaccess}. Next, we briefly present three centralized solutions: 1) Round Robin scheduling, 2) WiFresh from \cite{kadota2020age} and 3) Maximum Error First from \cite{ayan2019aoi}. In subsection \ref{subsubsec:polmef}, we are going to introduce a new protocol that combines the core ideas from \ref{subsubsec:polling} and \ref{subsubsec:mef}, hence consolidates the strong sides of both methods. As we are going to show in section \ref{sec:results}, our solution is able to outperform the other methods concerning control performance.
	
	\subsection{Contention-based Protocols}
	\label{subsec:randomaccess}
	\subsubsection{\textbf{ALOHA} }
	In our experimental framework, the simplest MAC protocol that we have implemented is the pure ALOHA proposed in \cite{abramson1970aloha}. It is based on the simple idea of transmitting any incoming data packet when it is ready to send. We know from the basics of wireless communications that this results in high packet loss if network traffic load is high.
	\subsubsection{\textbf{Slotted ALOHA (SA)}}
	Slotted ALOHA, which has originally been proposed in \cite{lawrence1975aloha}, is based on the idea that time is divided into equally long time slots and each user transmits with a constant channel access probability (CAP) $p_i$ when a slot begins or backs off with $1 - p_i$ probability. 
	
	SA has recently been studied in the context of AoI in \cite{chen2020age, han2020software}\footnote{In \cite{chen2020age, han2020software} authors refer to slotted ALOHA as ``Age-independent random access (AIRA)''.}. In particular, in \cite{han2020software}, the authors use SDRs programmed with GNU Radio similar to this work. As derived in \cite{chen2020age}, by using SA each loop achieves a mean AoI $\bar{\Delta}_{SA}$ given as:
	\begin{equation}
	\bar{\Delta}_{SA} = \frac{1}{p(1 - p)^{N - 1}},
	\label{eq:AoISA}
	\end{equation}
	where $N \geq 3$ is the number of users in the network. As proven in the same work, the age-optimal CAP $p^*$ for SA is given as $p_i^* = p^* = \slfrac{1}{N}, \, \forall i$. Throughout the paper, we assume that when SA is employed, the optimal channel access probability is selected. 
	
	For \eqref{eq:AoISA} to be valid, the nodes should always transmit the most recent state information. By adopting the LCFS queuing discipline, we make sure that this requirement is fulfilled. Moreover, the slot frequency and the frequency of the aging process should coincide which is the case for our work. 
	
	Time synchronization among SDRs, which is necessary for time slotted protocols such as SA, is realized through periodic transmission of beacon packets. Further details for synchronization are given in section \ref{subsec:synchronization}. 
	
	\subsubsection{\textbf{Age-dependent Random Access Protocol (ADRA)}}
	The ADRA protocol was proposed in \cite{chen2020age} by Chen et al. as an optimized age-dependent stationary randomized MAC policy for large-scale networks. It is a threshold-based policy in which each user accesses the shared medium with a predetermined CAP $p = p_i, \, \forall i$ only if its instantaneous AoI is not below a certain threshold value $\delta_i = \delta, \, \forall i$, i.e.:
	\begin{equation}
	p_i[t] = \begin{cases} 
	0 &\mbox{, if } \Delta_i[t] < \delta \\
	p & \mbox{, if }\Delta_i[t] \geq \delta \end{cases} 
	\label{eq:adrathreshold}
	\end{equation}
	\begin{figure}[t]
		\centering
		\includegraphics[width=.95\columnwidth]{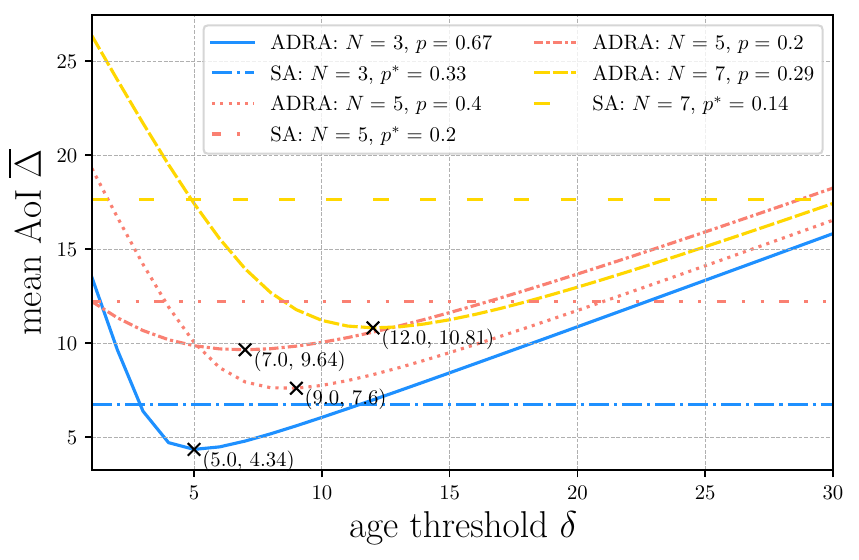}
		\caption{Network-wide mean AoI $\bar{\Delta}$ is plotted against age-threshold $\delta$ for selected number of users, $N = \{3, 5, 7\}$. $p$ denotes the channel access probability for the ADRA protocol. The horizontal lines show the minimum achievable AoI for slotted ALOHA with the age-optimal CAP $p^* = \slfrac{1}{N}$.}
		\label{fig:adra}
	\end{figure}

	Evidently from \eqref{eq:adrathreshold}, each source SDR $\sdr_i$ needs to know the instantaneous AoI at the receiver in order to decide whether it is eligible for data transmission. However, as the sensors do not have the perfect knowledge of the reception history at the receiver, the instantaneous AoI has to be estimated remotely. To overcome this issue, which is not directly addressed in theoretical works, our framework makes use of acknowledgment (ACK) packets transmitted by the GW device upon a successful reception. The instantaneous AoI estimation at the sensors is based on the assumption that every unacknowledged packet is lost on the sensor-to-GW link. In case of an unreliable control channel with a high loss ratio of the ACK packets or when they arrive with a significant delay, the ADRA protocol would overestimate the AoI at the receiver, thus leading to more frequent and redundant transmissions, hence increased network load.
	
	In their work, the authors derive the network-wide mean AoI for ADRA protocol as:
	\begin{equation}
	\bar{\Delta}_{ADRA} = \frac{\delta}{2} + \frac{1}{pq} - \frac{\delta}{2(\delta p q + 1 - p q)},
	\end{equation}
	with the successful status update probability $q$. To obtain the value for $q$ we refer to the original paper. Moreover, the optimal values for $\delta$ and $p$ can be obtained numerically. As suggested by the authors, we used the bisection method to find the optimal $\delta^*$ and $p^*$ values. When comparing the ADRA protocol to others, like in the SA case, we have used the optimal values for $\delta$ and $p$ during our measurements.  Fig. \ref{fig:adra} depicts the network-wide mean AoI $\bar{\Delta}_{ADRA}$ and $\bar{\Delta}_{SA}$ for various number of users $N = \{3, 5, 7\}$ and varying $\delta$ up to $30$. It is evident from the figure that when the right configuration is selected, the ADRA protocol outperforms the slotted ALOHA w.r.t. the mean AoI.
	\subsection{Contention-free Protocols}
	\label{subsec:contentionfreeprotocols}
	\subsubsection{\textbf{Round Robin (RR)}}
	The RR is a well-known scheduling policy from the literature that prioritizes each user one after another in a fixed order. Therefore, it is neither a channel-aware nor an application-aware scheduling algorithm. In our RR implementation, the users are prioritized in the same order as their unique control loop ID $i$. Given that at any time slot $t \in \mathbb{N}^+$ only a single source node $i \in \{1, 2, \dots, N\}$ is scheduled, the next node to schedule can be obtained by the simple rule:
	\begin{equation}
	i^*[t] = \arg \min_i ~ \{ t + N - i \mod N\},
	\label{eq:roundrobin}
	\end{equation}
	with the modulo operator$\mod$. There is always a single user $i$ that makes $t + N - i \mod N = 0$, where $i^*[1] = 1$, $i^*[2] = 2$, etc.
	In our framework, we enforce synchronization among users with the help of beacon packets as in SA and ADRA protocols. Therefore, each source node $i$ can track the current time slot index $t$ and thus detect the next allocated slot by \eqref{eq:roundrobin}.
	
	With constant number of users, RR results in periodical prioritization of every user. That is, every source node is scheduled once in every $N$ slots. If the destination can successfully decode all transmitted updates, the discrete time AoI of each user experiences a decrease from $N$ to $1$ with a periodicity of $N$ slots. In other words, every time when AoI reaches $\Delta_i[t'] = N$, it is followed by a reset to $\Delta_i[t' + 1] = 1$ in the subsequent slot. As a result, the long-term mean AoI of each source node is equivalent to its  mean AoI over a period of $N$ slots, which can be derived as a sum of arithmetic sequence as follows:
	\begin{equation}
	\bar\Delta_{RR} = \frac{1}{N}\bigl(\frac{N}{2}(1+N)\bigr) = \frac{N+1}{2}.
 	\label{eq:roundrobin_aoi}
	\end{equation}
	
	In spite of its simple operation, RR comes with some drawbacks in practical deployment. In addition to its dependence on time synchronization, the RR may cause underutilization of the network resources. In particular, the RR allocates certain amount of resource units, e.g., a time slot, exclusively to a user. This implies that if a transmission takes shorter than the allocated slot, the remaining portion of the resource is wasted. Especially, in typical connected robotics and remote monitoring scenarios, in which the sensors store a single packet of small size, finding the right slot duration to accommodate exactly a single transmission becomes a challenge. Our results in section \ref{sec:results} reveal the performance loss originated by shorter transmissions than a slot duration. However, in this work we do not tackle the slot duration adaptation problem and choose a fixed length throughout our measurements.
	
	\subsubsection{\textbf{WiFresh}}
	\label{subsubsec:polling}
	One of the most prominent examples of practical AoI research is WiFresh \cite{kadota2020age}, which is a polling-based protocol. Similar to this work, the authors consider multiple sources transmitting via SDRs to a base station (BS). The BS tracks the AoI of each source process and asks for a status update packet by sending a poll request. Additionally, it estimates the channel reliability $r_{i}^{ ch}(t)$ between a source device $i$ and the BS by the following equation: 
	\begin{equation}
	r_{i}^{ch}(t) = \frac{RX_{i}^{D}(t) + 1}{TX_{i}^{P}(t) + 1},
	\label{eq:channelrel}
	\end{equation}
	where $RX_{i}^{D}(t)$ and $TX_{i}^P(t)$ denote the number of successfully received data packets and transmitted poll packets in the last $0.5$ seconds, respectively. The next source node to poll is then determined by the max-weight policy as:
	\begin{equation}
	i^*(t) = \arg \max_i ~ \{r_i^{ch}(t) \tilde{\Delta}_i(t)\},
	\label{eq:wifresh}
	\end{equation}
	with $\tilde{\Delta}_i(t)$ being the estimated age of the freshest information about source node $i$. The necessity for the AoI estimation arises due to the fact that the source and destination nodes are not co-located and the sampling instances of the source nodes are unknown to the GW. Therefore, the GW is obliged to estimate the AoI remotely by tracking the elapsed sampling periods since the latest reception. Our approach is similar to the one considered in \cite{kadota2020age}. We would like to mention that this is an example challenge of system research that is revealed only prior to deployment and may be hidden for purely theoretical works.
	
	Moreover, please notice the round brackets we have used for the variables in \eqref{eq:wifresh}, which has the following reason: WiFresh is a polling-based protocol that operates asynchronous to the time slotted model we have introduced before. In other words, the GW does not have any notion of a network time slot and therefore immediately begins with the next polling procedure once the outstanding poll packet has been responded to by a data packet.
	
	The channel-awareness of WiFresh makes it suitable for environments where the nodes are highly mobile, thus experience time-varying link quality. In addition, as it does not rely on synchronization among users, one can argue for its lower complexity when compared to SA or ADRA. It is clear that in contrast to random access protocols, the packet success ratio is expected to be much higher as simultaneous channel access is avoided by virtue of the centralized polling mechanism.
	
	\subsubsection{\textbf{Maximum Error First Scheduler (MEF)}}
	\label{subsubsec:mef}
	The MEF scheduler has been proposed in \cite{ayan2019aoi} for two-hop cellular networks where the users are feedback control loops. \cite{ayan2019aoi} suggests to employ the MSE from \eqref{eq:MSEage} as the scheduling metric in a time-slotted resource allocation problem. As a result, at each time slot $t$, the next user to schedule is determined as:
	\begin{equation}
	i^*[t] = \arg \max_i ~ \{MSE_i[t]\}.
	\label{eq:mef}
	\end{equation}
	The MEF scheduler is an example of control-aware scheduling policies for wireless NCS that has only been studied in the context of theoretical research. To the best of our knowledge, there has not been any previous work that implements the algorithm in a practical setup. Therefore, the following design choices have been made in order to implement the MEF scheduler in our framework:
	\begin{itemize}
		\item The GW broadcasts a beacon packet every $20$ time slots that contains the transmission schedule during that period.
		\item Only the source node $i^*[t]$ that has been scheduled for transmission at time slot $t$ accesses the channel.
		\item The GW neglects the probability of a packet loss and allocates each of those $20$ slots in advance as if all transmissions were to be successful.
	\end{itemize} 
	We provide more details on beacon packets and time synchronization later in subsection \ref{subsec:synchronization}.
	\subsubsection{\textbf{Our Proposed Polling-based MEF Scheduler (pMEF)}}
	\label{subsubsec:polmef}
	The key difference of MEF from WiFresh is that MEF considers control system dependent parameters implicitly through MSE. On the other hand, WiFresh does not operate in a slotted fashion in contrast to MEF scheduler. This feature allows WiFresh to reduce the amount of idle time between two consecutive transmissions if the response to a poll packet comes earlier than the beginning of the next slot\footnote{In section \ref{sec:results}, we discuss the effect of this property of polling on the AoI and control performances in detail.}. Moreover, the MEF scheduler was originally proposed as a channel-unaware scheduling policy as it is evident from \eqref{eq:mef}. Therefore, we propose to combine the strengths of both schedulers in a polling-based, channel- and control-aware scheduler, that determines the next source node to schedule as:
	\begin{equation}
	i^*(t) = \arg \max_i ~ \{r_i^{ch}(t) MSE_i(t)\},
	\label{eq:polmef}
	\end{equation}
	with $r_i^{ch}(t)$ as in \eqref{eq:channelrel}. The MSE is obtained by substituting the instantaneous estimated AoI $\tilde{\Delta}_i(t)$ into \eqref{eq:MSEage}. As in the WiFresh case, the round brackets are used to emphasize asynchronous operation of the scheduler to sampling process in contrast to time-slotted implementations such as MEF or RR.

	\subsection{A new metric for control-aware scheduling: the normalized MSE (nMSE)}
	The MSE, as defined in \eqref{eq:MSEage}, has been used for control-aware scheduling in previous works \cite{ayan2019aoi, ayan2021age}. However, by definition it is strongly system-dependent, hence its unit varies from one control application to another. As a result, when making scheduling decisions that consider system parameters, as in the case of MEF scheduler or our proposed pMEF scheduler, one can not employ the MSE in its raw form in systems design. In other words, it may not capture the urgency of transmission for different applications. More precisely, the scheduling decision based on raw MSE would correspond to the comparison of multiple numbers in different units and orders of magnitude.
	
	As a solution to this problem, we propose and employ the \textit{normalized mean-squared error (nMSE)} that is defined as:
	\begin{equation}
		\left\Vert MSE_i(t) \right \Vert \triangleq \dfrac{MSE_i(t)}{MSE_{\Delta_i = 1}}
		\label{eq:normalizedMSE}
	\end{equation}
	where $MSE_{\Delta_i = 1} \triangleq MSE_i(t) | _{\Delta_i(t) = 1}$. In simple words, we divide the MSE of each control sub-system by the MSE when the AoI is $1$. It is important to mention that the normalization factor, i.e., $MSE_{\Delta_i = 1}$, is equal to the trace of the covariance matrix, which is the only addend in the RHS of \eqref{eq:MSEage} when $\Delta_i(t) = 1$. Similar to the MSE, nMSE is zero when $\Delta_i(t) = 0$ and is strictly increasing with $\Delta_i$ since the denominator takes a positive value\footnote{The strictly increasing property of MSE had been shown in previous works, e.g., in \cite{ayan2019aoi}.}. 
	
	The normalized MSE can be seen as an adaptation of the MSE to the so called ``age-penalty" or ``non-linear aging" from the existing literature \cite{zheng2019closed, sun2017update}. The concept of non-linear aging has been proposed to represent the information losing its usefulness over time with a varying speed. In those works, the authors investigate well-known non-linear functions of AoI, such as $f(\Delta) = e^{a\Delta}$ and $f(\Delta) = \Delta^a$, with $a \geq 0$. In contrast to such system-independent penalty functions, the nMSE is a way of defining control-aware age-penalty functions as it depends on AoI, system matrix and the noise covariance matrix. Moreover, it captures the growth of the mean-squared estimation error relative to the value that it takes if the information has been generated in the previous sampling period. Through normalization, we are able to unify heterogeneous control applications in a dimensionless quantity. In section \ref{subsec:casestudy}, we present a case study utilizing the nMSE for wireless resource management, where the MSE is not directly applicable.

	By definition, the pMEF scheduler depends on the system dynamics as it utilizes the instantaneous nMSE, which is a normalized version of the MSE. More specifically, the centralized scheduler requires the knowledge of the system dependent parameters $\bm{A}_i$ and $\bm{\Sigma_i}$ to be able to obtain $MSE_i(t)$ given the AoI as in \eqref{eq:MSEage}. Nevertheless, as those parameters are time-invariant, due to the fact that we are dealing with LTI systems, a single information exchange prior to operation is sufficient.
	
	\section{Design and Implementation}
	\label{sec:implementation}
	\subsection{Hardware and Software}
	\label{subsec:hwsw}
	Our experimental setup consists of $N \in \{2, 3, \dots, 15\}$ plant processes programmed in Python programming language. Each plant process $\plant_i$ generates periodic packets that are forwarded to $\sdr_i$ using a UDP socket\footnote{Each plant process $\plant_i$ and SDR $\sdr_i$ run on the same machine.}. Once the packet is received by the SDR, it traverses through multiple packet processing blocks programmed in C++ with GNU Radio.
	
\begin{figure}[t]
		\centering
		\includegraphics[width=0.8\columnwidth]{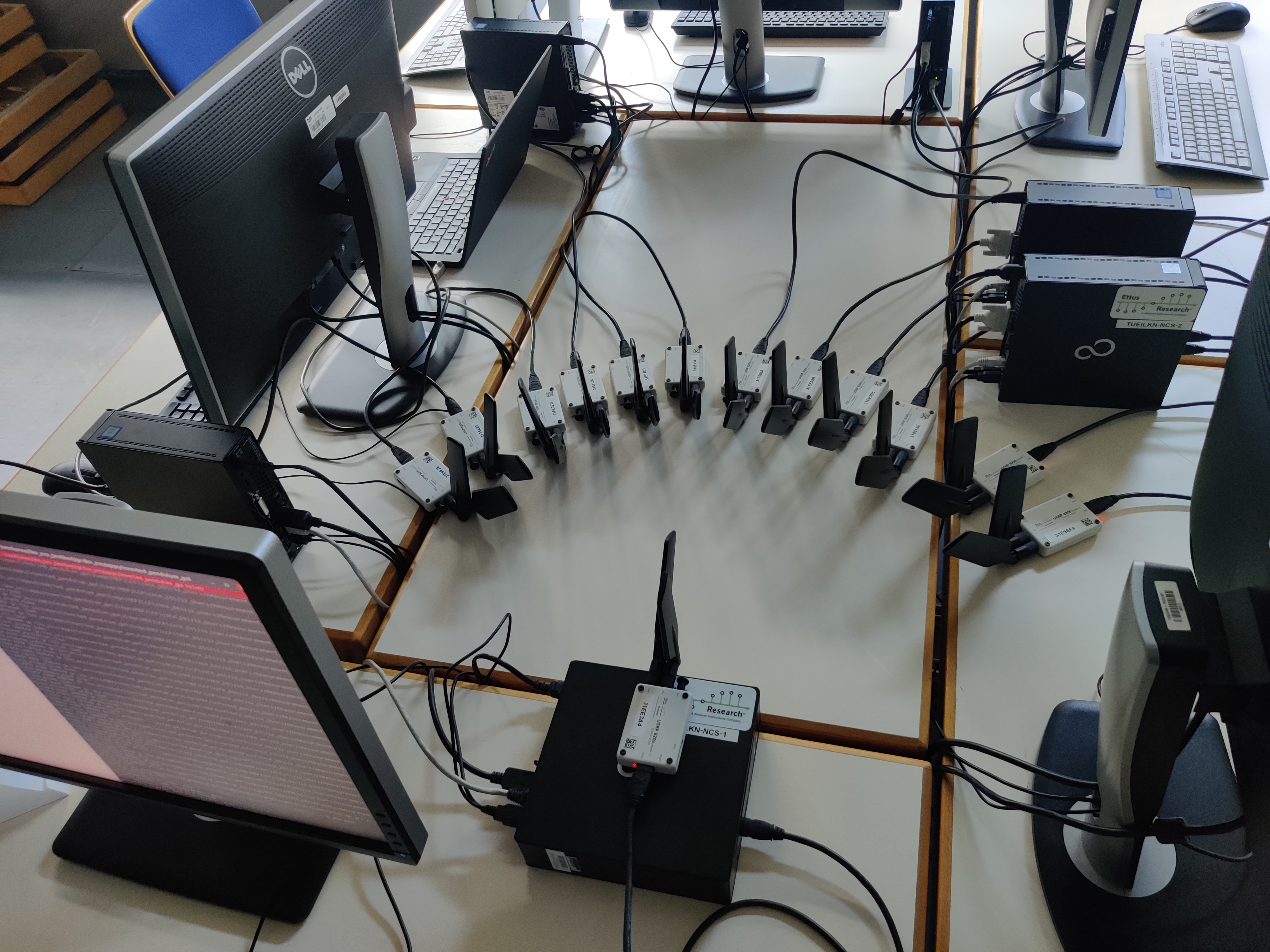}
		\caption{A photo of our testbed while taking measurements with $12$ control sub-systems. 
		}
		\label{fig:testbed}
	\end{figure}

	Our testbed is composed of $8$ computers running Ubuntu 20.04.3 LTS operating system. Ettus Research's USRP\textsuperscript{TM} B200mini-i and B205mini-i SDRs are used as the source and destination for wireless data transmission. Fig. \ref{fig:testbed} shows a photo of our experimental testbed while measurements with $12$ control sub-systems were being conducted. Note that there are $12$ SDRs responsible for the transmission of status update packets and an additional SDR serving as GW. In contrast to \cite{han2020software}, we have not directed the data flow of multiple source processes into a single SDR.
	
	In our framework, we have a clear separation of the application layer and the wireless communication stack. Specifically, the status update packets are generated and written to a local UDP socket that is read by the GNU Radio signal processing blocks. The wireless network behind the UDP socket is completely transparent to the application, i.e., the control system. Similarly at the GW, the interfacing between the GNU Radio process and the controller processes is done by employing local UDP sockets. By choosing a clear separation between the wireless networking stack and the application layer, we aim to simplify the integration of any internet protocol based application into our framework, hence to remove the barrier to its adoption.
	
	An automation script is used to reduce the influence of a human operator on the results when the measurements are started, repeated and stopped. Additionally, we ignore the first and last $5 \,s$ of each $30 \,s$ long measurement run in the data collection to avoid transitional effects of the startup and completion phases. 
	
	\subsection{Synchronization}
	\label{subsec:synchronization}

	Time synchronization is necessary in order to realize the time-slotted MAC protocols introduced in Sec. \ref{sec:protocols} such as SA and ADRA random access protocols or RR scheduling. To that end, we follow a similar approach as in \cite{han2020software} and employ periodic transmission of beacon packets at the beginning of each $20$ slots long \textit{frame} structure. A beacon packet is composed of three main fields: 
	\begin{itemize}
		\item \textbf{MAC header}: Contains information such as packet type, MAC sequence number, source and destination addresses.
		\item \textbf{Payload}: Contains information specific to the employed MAC protocol, such as frame length, duration of a time slot, i.e., $T_s = 10$ ms, index of the time slot and the transmission schedule if applicable, e.g., for  MEF.
		\item \textbf{CRC}: Contains the 16-bits long cyclic redundancy check (CRC) field used to detect errors in the data reception, mainly caused by packet collisions in our setup.
	\end{itemize}
	Upon the detection of a beacon packet, each $\sdr_i$ marks the current time as the beginning of the next frame and sets the current slot to the time slot index contained within the \textit{Payload} field\footnote{Information on slot duration and frame length are contained in the beacon packet as well, although they are assumed constant in this work. The reason is to increase the flexibility of our implementation and facilitate the study on the effect of varying slot on network and control performance.}. This is based on the assumption that the difference in processing delays at each $\sdr_i$ is negligible. GNU Radio's \textit{high\_res\_timer} library has been used for time stamping purposes with high resolution.
	
	\section{Experimental Results and Evaluation}
	\label{sec:results}
	In order to simplify the implementation and the analysis of the results, we have selected scalar control loops of $3$ different classes. The least challenging category of systems are $\mathcal{I}_{easy} = \{1, 4, 7, 10, 13\}$ with the system matrix $\statemat_1 = \statemat_4 = \dots = \statemat_{13} = 1.0$. The second and third classes of systems, i.e., $\mathcal{I}_{mid} = \{2, 5, 8, 11, 14\}$ and $\mathcal{I}_{hard} = \{3, 6, 9, 12, 15\}$ have the system matrices $\statemat_{2} = \dots = \statemat_{14} = 1.1$ and $\statemat_{3} = \dots = \statemat_{15} = 1.2$, respectively. The relationship between the system matrix and difficulty of control can be  deducted from \eqref{eq:statespace} intuitively which shows the proportionality between the current state $\statevec_i[t]$ and the next state $\statevec_i[t+1]$. Additionally, the input and covariance matrices are chosen as $\inputmat_i = 1.0, \forall i$ and $\bm{\Sigma}_i = 1.0, \forall i$. 
	
	The design of the LQR controller has been done with $\bm{Q}_i = 100.0$ and $\bm{R}_i = 1.0$ for all control loops. In simple words, this means that the state error is penalized hundred times more than the control effort while calculating the optimal control input in the infinite horizon LQR problem. The optimal $\bm{L^*}_i$ is obtained from the solution of the discrete algebraic Riccati equation as given in \eqref{eq:riccatisolution}. 
	
	Our results are obtained by performing $20$ repetitions of $30$ seconds long measurement runs. As mentioned in \ref{subsec:hwsw}, we did not consider the first and last 5 seconds of each run to avoid transitional effects. Therefore, the evaluation of each metric starts after the $500$-th discrete time step and ends with the $2500$-th time step. As a result, the network-wide mean AoI is obtained as:
	\begin{equation}
	\bar{\Delta} = \frac{1}{2000 \cdot N}\sum_{t=501}^{2500} \sum_{i=1}^{N} \Delta_i[t]
	\label{eq:meanAoI}
	\end{equation}
	where each instantaneous AoI $\Delta_i[t]$ is measured at the end of each sampling period $t$. In order to capture the control quality, we have selected the mean squared estimation error $\overline{MSE}$ and the LQG cost $\overline{F}$ with:
	
	\begin{equation}
		\overline{F} \triangleq \frac{1}{2000 \cdot N}\sum_{i=1}^{N} \sum_{t=501}^{2500} (\boldsymbol{x}_i[t])^T \boldsymbol{Q}_i \boldsymbol{x}_i[t] +  (\boldsymbol{u}_i[t])^T \boldsymbol{R}_i \boldsymbol{u}_i[t]
		\label{eq:simLQG}
	\end{equation}
	with $\overline{F}_i$ as in \eqref{eq:lqg_cost}. The calculation of $\overline{MSE}$ is analogue to $\overline{\Delta}$ and can be obtained by replacing $\Delta_i[t]$ in \eqref{eq:meanAoI} with $MSE_i[t]$, i.e.:
	
	\begin{equation}
	\overline{MSE} = \frac{1}{2000 \cdot N}\sum_{t=501}^{2500} \sum_{i=1}^{N} MSE_i[t]
	\label{eq:simMSE}
	\end{equation}
	
	It is important to state that due to the selection of $\bm{\Sigma}_i = 1.0, \forall i$, the denominator in the RHS of \eqref{eq:normalizedMSE} becomes $1$. Therefore, the raw MSE and the normalized nMSE are equivalent for sections \ref{subsec:contentionbasedresults} and \ref{subsec:contentionfreeresults}, i.e., $\overline{MSE} = \overline{\left \Vert MSE \right \Vert}$. Note that this does not apply to Sec. $\ref{subsec:casestudy}$, in which we introduce a new control loop class into the network.
	
	\subsection{Contention-Based Protocols' Performance}
	\label{subsec:contentionbasedresults}
	\begin{figure}[t]
		\centering
		\includegraphics[width=.9\columnwidth]{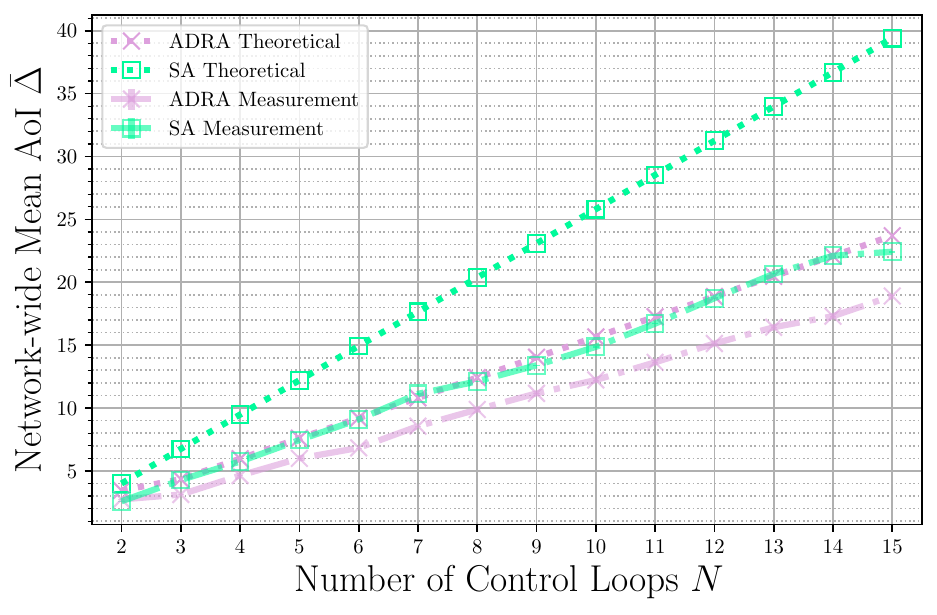}
		\caption{Mean AoI of contention-based access protocols, i.e., slotted ALOHA (SA) and age dependent random access (ADRA). Vertical bars illustrate 99\% confidence intervals.}
		\label{fig:ra_mean_aoi}
	\end{figure}
	
	In Sec. \ref{subsec:randomaccess} we have introduced three contention-based MAC protocols, namely ALOHA, SA and ADRA. Fig. \ref{fig:ra_mean_aoi} presents the measured mean AoI and its theoretical expectation, i.e., $\bar{\Delta}_{SA}$ and $\bar{\Delta}_{ADRA}$. We do not include ALOHA in the figure because of presentation purposes. That is, the ALOHA protocol performs significantly worse than the other two already with a very low number of users in the network. For instance, the instantaneous AoI up to $1900$ were observed in one of the measurements for $N=3$. Therefore, we omit ALOHA in the remaining evaluation since it is not suitable for time-sensitive wireless networks with multiple users.
	
	From the figure, we observe that the ADRA protocol outperforms the age-independent protocol SA as expected. However, we observe a deviation between the measurement results and the theoretical results from \cite{han2020software}. In fact, our framework is able to achieve better results than analytical mean values. In our opinion, this has two main reasons: 1) simultaneous transmissions are being decoded in spite of their overlapping. This issue has already been raised in \cite{han2020software}. 2) The transmission of a packet does not occupy a full slot. In our framework, a slot is $10$ ms long, whereas our measurements indicate an approximate transmission duration of $3$ ms for each data packet. As we do not force any synchronization in the application layer, this allows some of the packets to miss each other in time although they are transmitted in the same slot. This phenomenon increases the packet delivery rate per slot far beyond one, which causes an improvement over the theoretical expectation.
	
	\begin{figure}[t]
		\centering
		\includegraphics[width=.9\columnwidth]{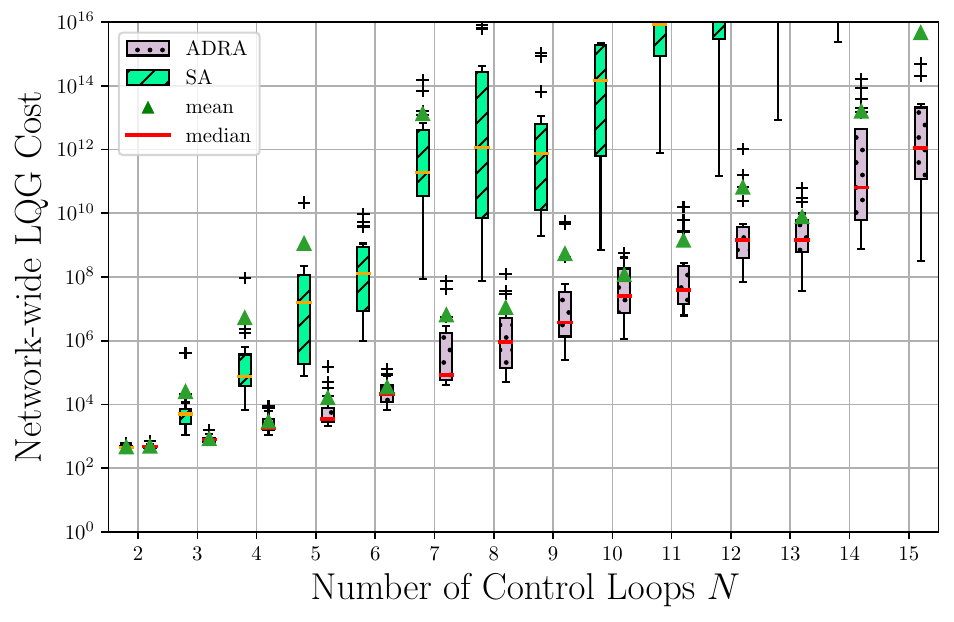}
		\caption{Control performance of contention-based protocols. It is captured by the LQG Cost $\overline{F}$ as defined in \eqref{eq:simLQG}, where a lower LQG cost represents a higher performance. y-axis has been limited for presentation purposes.}
		\label{fig:ra_lqg_cost}
	\end{figure}
	
	If we look at Fig. \ref{fig:ra_lqg_cost}, we observe that the LQG cost representing the control performance shows divergent behavior for both contention-based protocols. Especially, already for $N = 8$, SA reaches an LQG cost up to $10^{16}$ indicating an instability of the system state. The same applies to ADRA for $N=15$ showing the inadequacy of these protocols for multi-user scenarios with time-sensitive control applications.
	
	\subsection{Contention-Free Protocols' Performance}
	\label{subsec:contentionfreeresults}
	From the fundamentals of communications theory, we know that the main strength of the contention-free protocols over random access is their significantly lower packet loss rate. This comes at a price of increased complexity and communication overhead, as in the case of polling-based protocols. First, let us analyze the performance of contention-free protocols w.r.t. information freshness.
	
	\begin{figure}[t]
		\centering
		\includegraphics[width=\columnwidth]{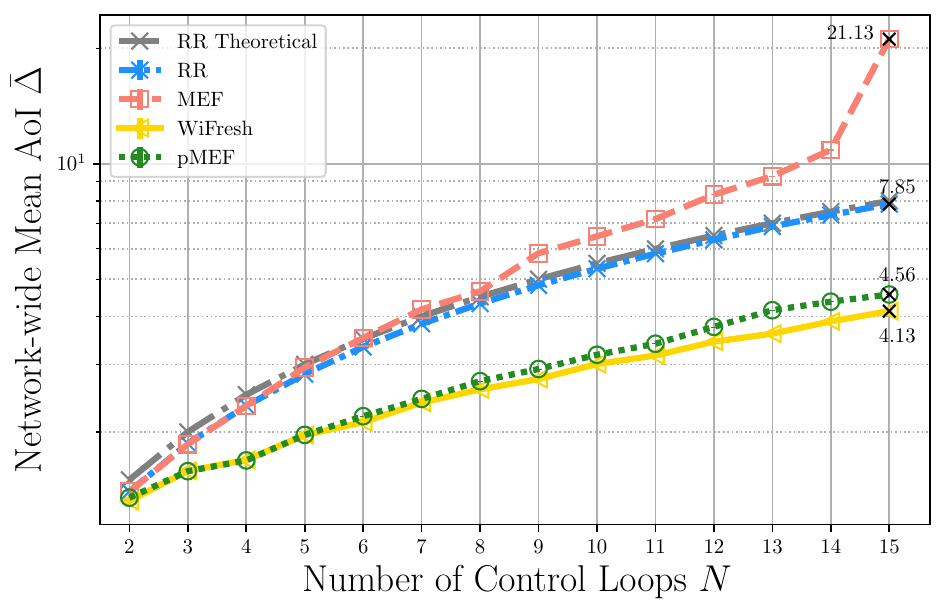}
		\caption{Mean AoI of contention-free protocols, i.e., round robin (RR), maximum error first (MEF), WiFresh and polling-based MEF (pMEF). Vertical bars illustrate 99\% confidence intervals. y-axis is drawn on logarithmic scale.}
		\label{fig:sch_mean_aoi}
	\end{figure}
	
	Fig. \ref{fig:sch_mean_aoi} presents the network-wide AoI for $2\leq N \leq 15$. We observe that the polling-based WiFresh protocol outperforms all others by at least $10\%$ as in the case of pMEF. This is an expected result due to the heterogeneous prioritization of sub-systems by the pMEF algorithm caused by its control-awareness. In other words, while WiFresh considers the AoI and therefore polls sub-systems in a round robin fashion under constant channel conditions, the pMEF allocates a bigger portion of the network resources to the class of more challenging sub-systems $\mathcal{I}_{hard}$. This leads to an unbalanced distribution of AoI in the network and increases $\bar{\Delta}$. Nevertheless, as we are going to show later in this section, pMEF is able to achieve better performance for the given control task via its ability to identify the most relevant information.
	
	In our setup, the average polling time, which is the time between a poll request and the reception of the corresponding data packet, is shorter than a time slot. Thus, the beacon-based protocols, i.e., RR and MEF, achieve lower throughput, less transmissions and hence higher AoI than WiFresh and pMEF. This leads to resource scarcity and longer idle periods for less critical sub-systems in the case of MEF. As a result, the gap in AoI between control-aware and control-unaware protocols, i.e., MEF and RR, is increased. Particularly, MEF achieves $\bar{\Delta}$ beyond $20$ for $N=15$ while RR achieves less than $10$ for the same number of sub-systems. Note that the experimental AoI for RR matches the theoretical mean AoI derived in \eqref{eq:roundrobin_aoi}.
	
	\begin{figure}[t]
		\centering
		\includegraphics[width=\columnwidth]{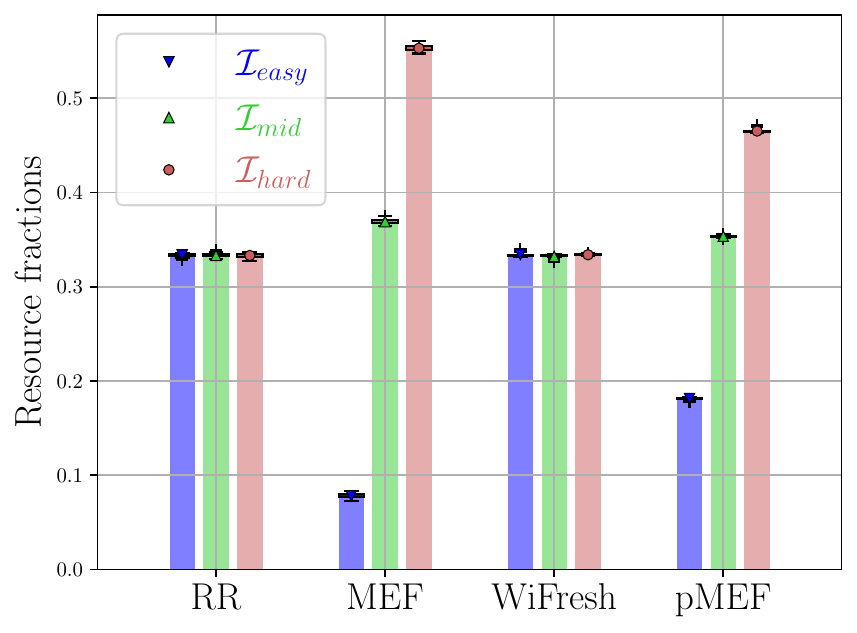}
		\caption{The fractions of network resources allocated to each control class. Control-unaware protocols. i.e., RR and WiFresh treat all system classes equally. On the other hand, the control-aware protocols, i.e., MEF and pMEF lead to an unbalanced distribution of resources.}
		\label{fig:fractions}
	\end{figure}
	As a next step, we present Fig. \ref{fig:fractions} which shows the fraction of network resources allocated to each control class $\mathcal{I}_{easy}$, $\mathcal{I}_{mid}$ and $\mathcal{I}_{hard}$. In fact, it confirms that all classes are treated equally when RR and WiFresh are applied, whereas MEF and pMEF schedule more challenging sub-systems more frequently.

\begin{figure}[t]
		\centering
		\includegraphics[width=\columnwidth]{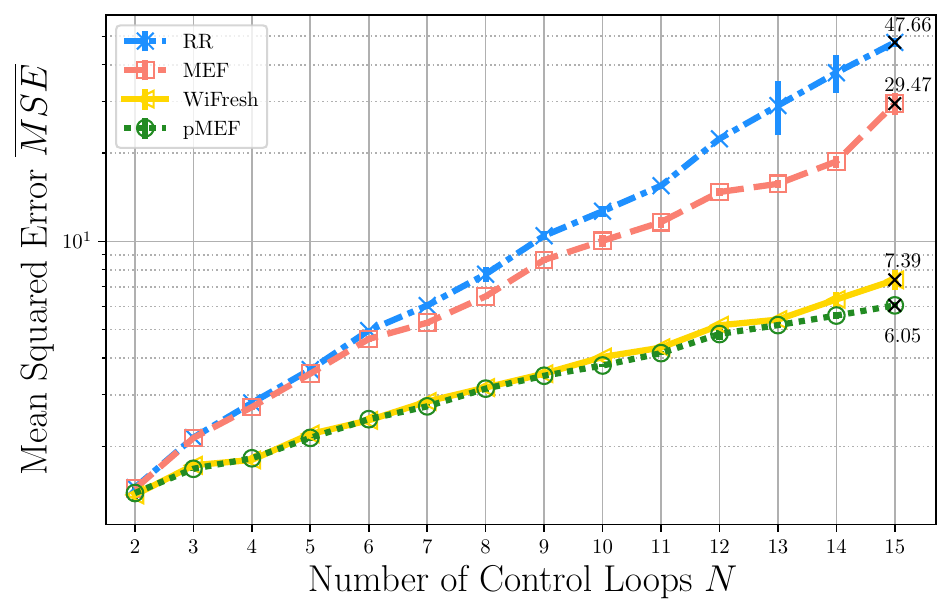}
		\caption{Estimation performance of contention-free protocols, i.e., RR, MEF, WiFresh and pMEF. It is captured by the MSE as defined in \eqref{eq:simMSE}. Note that a lower $\overline{MSE}$ represents a higher performance. Vertical bars illustrate 99\% confidence intervals. y-axis is drawn on logarithmic scale.}
		\label{fig:sch_mse}
	\end{figure}

	So far we have only presented the results of contention-based protocols concerning information freshness. However, as we are dealing with NCSs that are communicating in order to achieve a certain control goal, we need to go beyond AoI and focus on control-related KPIs such as MSE and control cost. Firstly, we present Fig. \ref{fig:sch_mse} that shows the estimation performance in the network captured by  $\overline{MSE}$. It is evident from the figure that the control-aware protocols, i.e., MEF and pMEF, outperform their direct competitors, i.e., RR and WiFresh, respectively. Especially, as the resource scarcity becomes more significant, e.g., $N=15$, the importance of control-awareness stands out. That is, MEF is able to achieve relatively lower $\overline{MSE}$ than RR, although it performs worse than RR w.r.t $\overline{\Delta}$. One can also say that the information freshness is traded for an increase in estimation performance. A similar behavior is observed when pMEF and WiFresh are compared, i.e., pMEF outperforms WiFresh by up to $18\%$ when there are $15$ control sub-systems in the network.
	
	\begin{figure}[t]
	\centering
	\includegraphics[width=\columnwidth]{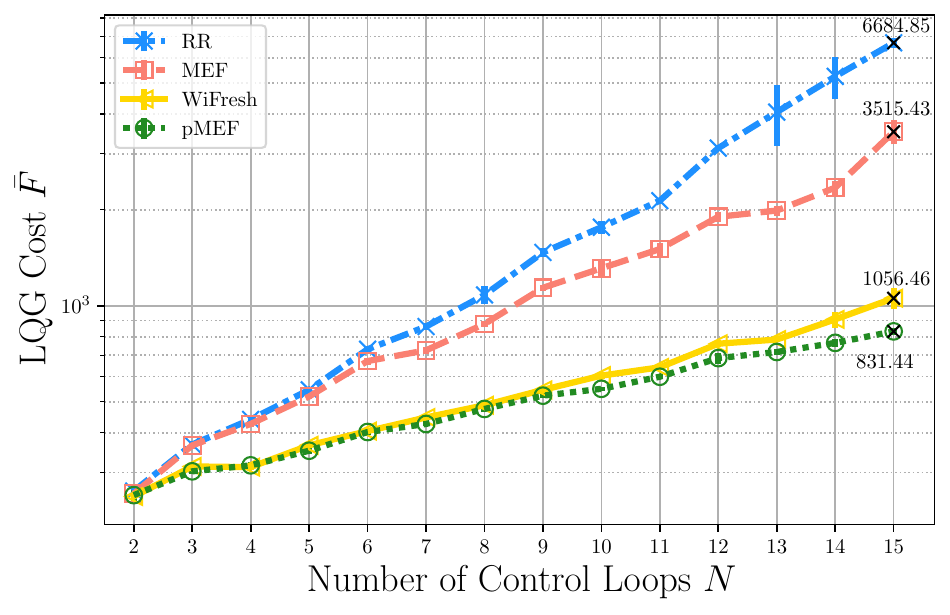}
	\caption{Control performance of contention-free protocols, i.e., RR, MEF, WiFresh and pMEF. It is captured by the LQG Cost $\overline{F}$ as defined in \eqref{eq:simLQG}. Note that a smaller $\overline{F}$ represents a higher performance. Vertical bars illustrate 99\% confidence intervals. y-axis is drawn on logarithmic scale.}
	\label{fig:sch_lqg_cost}
\end{figure}
	
	The MSE captures how accurate the remote system state is estimated at the monitoring process. On the other hand, the quality of control is not measured by the estimation accuracy but rather by the state error and the control effort that is spent in order to drive the state to the desired set point. However, the control performance is strongly intertwined with the estimation accuracy, as discussed in details in section \ref{subsec:estimationerror}. Due to this indirect relationship between the estimation and control performances, we observe a similar trend for the LQG cost as for MSE. Fig. \ref{fig:sch_lqg_cost} presents the main results of this work, i.e., the network-wide control cost $\overline{F}$ for different protocols. From the figure, we can see that pMEF is able to outperform the WiFresh protocol by up to $21\%$. The beacon-based protocols' performance follows a similar trend, with MEF outperforming RR by $47\%$. Furthermore, please notice that the contention-free protocols clearly outperform the contention-based schemes in terms of LQG cost.
	
\subsection{A Real-Life Application Case Study: Inverted Pendulum}
\label{subsec:casestudy}

\begin{figure}[t]
	\centering
	\includegraphics[width=.4\columnwidth]{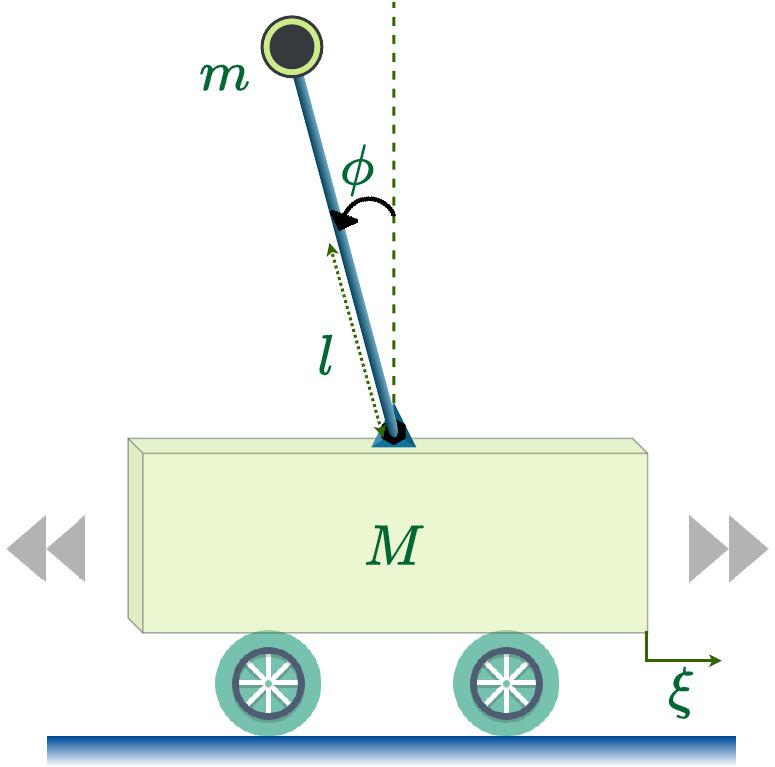}
	\caption{An inverted pendulum with motorized cart.}
	\label{fig:invertedpendulum}
\end{figure}

In the previous sections, we have shown the performance improvement of our proposed scheduling algorithm, over the existing protocols from the SotA. The selected scalar control systems were theoretical ones to illustrate this effect in a simple scenario. In this section, we introduce the emulation of a real-life application to our network, the inverted pendulum (IP). IP is a well-studied control application that is widely used in control theory textbooks \cite{astrom2008feedback}. As depicted in Fig. \ref{fig:invertedpendulum}, it consists of a pendulum mounted on a motorized cart where the controller's objective is to hold the pendulum in an upright position by moving the cart back and forward. For the sake of completeness, we provide the two continuous-time equations of motion around the unstable upward equilibrium:
\begin{align}
	&(I + m l^2) \ddot{\phi} - m g l \phi = m l \ddot{\xi}, \\ 
	&(M + m) \ddot{\xi} + b\dot{\xi} - m l \ddot{\phi} = u,
\end{align}
for the state vector $\bm{x} = [\xi, \dot{\xi}, \phi, \dot{\phi}]^T$. Here, $\xi$ is the position of the cart, $\phi$ is the deviation of the pendulum's position from equilibrium, $u$ is the input force applied to the cart. $M$ and $m$ are the mass of the cart, mass of the pendulum, respectively. $l$ denotes the length to pendulum's center of mass. In addition, $b$ is the coefficient of friction for the cart and $I$ is the moment of inertia of the pendulum. $g$ is the standard acceleration due to gravity. The selected set of parameters are summarized in the following table:
\begin{center}
	\begin{tabular}{ l | c }
		$M$ & $0.5$ kg  \\ \hline
		$m$ & $0.2$ kg \\ \hline
		$b$ & $0.1$ N/m/s  \\ \hline
		$l$ & $0.3$ m \\ \hline
		$I$ & $0.006$ kg$\text{m}^2$ \\ \hline
		$g$ & $9.81$ m/$\text{s}^2$
	\end{tabular}
\end{center}

As we are working with digital systems, we are interested in the discrete-time state-space representation of the form \eqref{eq:statespace}. Therefore, we select a sampling frequency of the system as $100$ Hz that leads to the following state and input matrices:
\begin{equation*}
	\bm{\tilde{A}} = \begin{bmatrix}
	1 & 0.01  & 0.0001 & 0 \\
	0 & 0.9983 & 0.0191 & 0.0001 \\
	0 & 0 & 1.0017 & 0.01 \\
	0 & \text{-}0.0049 & 0.3351 & 1.0017
	\end{bmatrix},
\bm{\tilde{B}} = \begin{bmatrix}
0.0001 \\
0.0182 \\
0.0002 \\
0.0454 
\end{bmatrix}.
\end{equation*}
Moreover, the noise covariance matrix is selected as:
\begin{equation*}
\bm{\tilde{\Sigma}} = \begin{bmatrix}
6.4 \cdot 10^{-7} & 0 & 0 & 0 \\
0 & 4.9 \cdot 10^{-7} & 0 & 0 \\
0 & 0 & 2.742 \cdot 10^{-5} & 0 \\
0 & 0 & 0 & 4.874 \cdot 10^{-5}
\end{bmatrix}.
\end{equation*}

The LQR method is used to determine the stabilizing feedback gain with weighting matrices $\bm{Q} = \text{diag}(5000, 0, 100, 0)$ and $\bm{R} = 1$ as in \eqref{eq:lqg_cost}.

In order to see the proposed nMSE metric in action, we repeat our measurements with $15$ control sub-systems, where we substitute all the sub-systems of class $\mathcal{I}_{mid}$ with IPs, i.e., $\bm{A}_2 = \bm{A}_5 = \dots = \bm{A}_{14} = \bm{\tilde{A}}$. We modify the input and noise covariance matrix analogously, as described above.

\begin{figure}[t]
\centering
\includegraphics[width=\columnwidth]{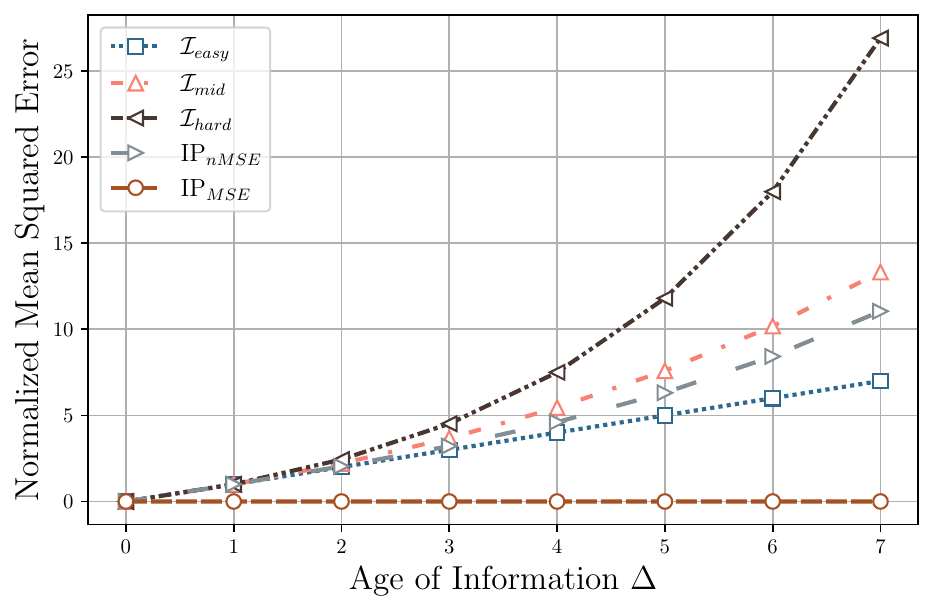}
\caption{The normalized mean squared error plotted against AoI, $\Delta$ for different control sub-systems used in this work, i.e., inverted pendulum (IP), $\mathcal{I}_{easy}$, $\mathcal{I}_{mid}$, $\mathcal{I}_{hard}$ as before. In addition, we present the raw MSE for IP before the normalization step from \eqref{eq:normalizedMSE} to illustrate its necessity.}
\label{fig:norm_mse}
\end{figure}

Fig. \ref{fig:norm_mse} shows the evolution of nMSE with increasing AoI together with the MSE for IP without the normalization step from \eqref{eq:normalizedMSE}. It illustrates the different growth speed of the nMSE for our considered application classes. Additionally, it reveals that the IP lies between the $\mathcal{I}_{hard}$ and $\mathcal{I}_{easy}$ classes with respect to the nMSE. Note that due to the significant difference in magnitude of order between $\text{IP}_{MSE}$ and other curves, the usage of the raw MSE would lead to resource starvation for sub-systems of class IP and destabilization of the corresponding control loops. Therefore, the following discussion considers only the usage of nMSE both for MEF and pMEF strategies.

Similar to the previous subsections, we measure the control KPIs in order to validate the applicability of our proposed protocol for real-life applications. To that end, we have recorded pendulum angle $\phi$ and cart position $\xi$ trajectories throughout $20$ measurements. In order to narrow down the focus on IP, the following discussion is limited to the IP relevant metrics such as $\phi$ and $\xi$ and does not contain the detailed state trajectories of other sub-systems of class $\mathcal{I}_{easy}$, i.e., $i \in \{1, 4, 7, 10, 13\}$ and of class $\mathcal{I}_{hard}$, i.e., $i \in \{ 3, 6, 9, 12, 15\}$.

\begin{figure}[t]
	\centering
	\includegraphics[width=\columnwidth]{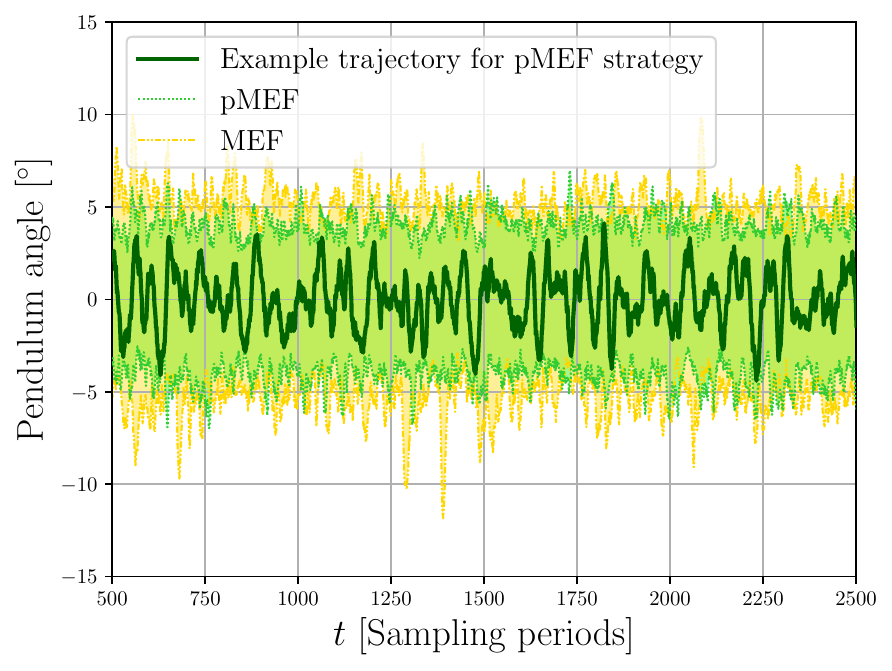}
	\caption{An example trajectory of the pendulum angle $\phi$ when maximum error first (MEF) and polling MEF schedulers are applied. $\phi$ is plotted in degrees.}
	\label{fig:angle_traj}
\end{figure}

\begin{figure}[t]
	\centering
	\includegraphics[width=\columnwidth]{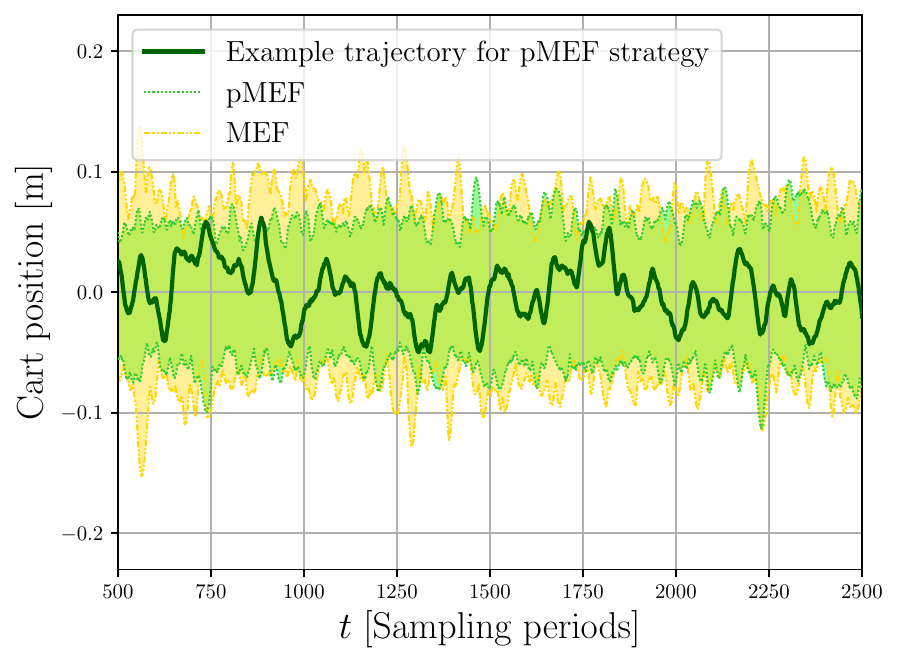}
	\caption{An example trajectory of the cart position $\xi$ when maximum error first (MEF) and polling MEF schedulers are applied. $\xi$ is plotted in meters.}
	\label{fig:pos_traj}
\end{figure}

In Fig. \ref{fig:angle_traj}, we present an example trajectory of $\phi_i[t]$ in degrees for $t \in [500, 2500]$ and a randomly selected loop $i$. It has been recorded during one of the measurements when pMEF scheduler operating with nMSE was in use\footnote{The selection of the specific measurement run and loop have been done in a random fashion and they do not represent an outlier w.r.t. control performance.}. From the figure, we are able to observe that the pendulum angle is kept within $\pm$ 5 degrees. In addition, Fig. \ref{fig:angle_traj} shows the maximum and minimum values that are reached by all IPs in the network when MEF and pMEF is employed. Due to the higher sensor-to-controller delivery rate of pMEF compared to MEF, the pMEF achieves a better control performance w.r.t. $\phi$. The same conclusion can be drawn if we look at Fig. \ref{fig:pos_traj} where the minimum and maximum $\xi$ values are presented. In particular, we are able to observe larger spikes of $\xi$ achieved by MEF than pMEF throughout the measurements. To put it another way, the cart needed to move further away from its desired set point, i.e., $\xi = 0$ in order to keep the pendulum upright.

\begin{figure}[t]
	\centering
	\includegraphics[width=\columnwidth]{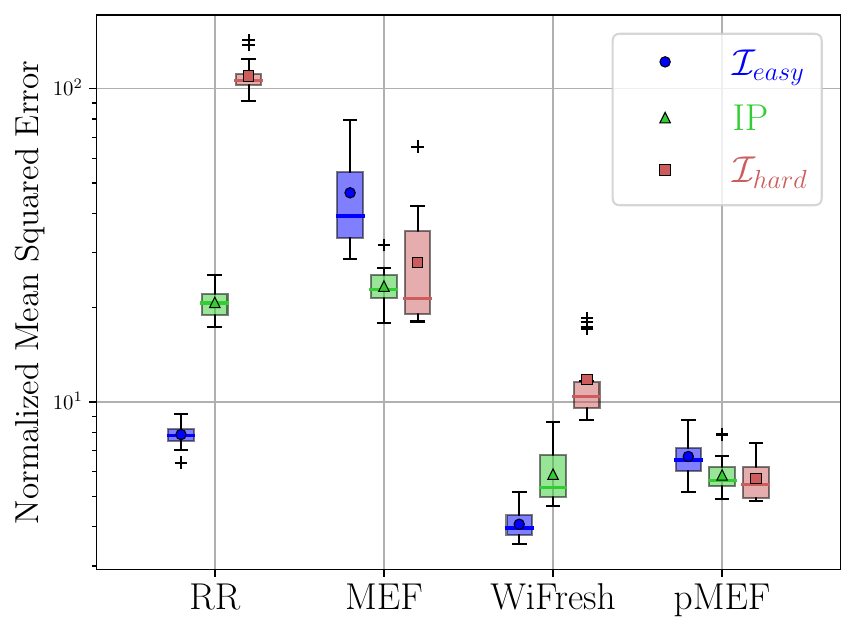}
	\caption{The normalized mean squared error (nMSE) achieved when round robin (RR), maximum error first (MEF), WiFresh and polling MEF schedulers are employed. y-axis is drawn on logarithmic scale.}
	\label{fig:mses_pend}
\end{figure}

Last but not least, Fig. \ref{fig:mses_pend} depicts the average nMSE achieved when RR, MEF, WiFresh and pMEF are used. Each boxplot represents a control class-scheduling strategy combination. In other words, it presents the contribution of each control class to the overall nMSE performance separately. From the figure, we can see that control-unaware strategies, namely the RR and WiFresh strategies lead to an increased nMSE for the $\mathcal{I}_{hard}$ class systems. This is an expected result of equal treatment of all sub-systems in the network which lead to higher error values for more critical applications. On the other hand, as we know from Fig. \ref{fig:fractions}, MEF and pMEF allocate more resources to $\mathcal{I}_{hard}$ systems than IP and $\mathcal{I}_{easy}$. As a result, they are able to balance out the higher task criticality of those sub-systems through their awareness of nMSE displayed in Fig. \ref{fig:norm_mse}.

\section{Conclusion and Final Remarks}
\label{sec:conclusion}

AoI has been used for remote monitoring and control scenarios to quantify information freshness. However, in a network of heterogeneous control applications, providing freshness may not guarantee optimal performance due to diversified system dynamics and task-criticalities. Hence, customization of network through application dependent metrics is beneficial for satisfying heterogeneous demands of such systems.

In this work, we study practical implementation of various customized MAC protocols that have been proposed for increased information freshness and control performance. In addition, we propose and implement a new task-oriented contention-free protocol that considers the quality of estimation at the monitor and employs the estimation error as utility for resource scheduling. Through real-world measurements using SDRs, we show that our proposed solution outperforms the selected existing strategies w.r.t. control and estimation performance. Moreover, we propose a new metric called the normalized mean squared error that is a modified version of the previously proposed age-dependent MSE for NCSs. We demonstrate its applicability as a scheduling metric when the control loops are of heterogeneous type and dimensions.
Our results reveal the high potential in cross-layer protocol design for task-oriented communications and networked control.

We expect the era of semantic communications to bring various research domains together implying a convergence of multiple layers of the communication stack. Particularly in task-oriented communications, lower layers are expected to be aware of the information content and track the semantics of data, such as AoI and value of information. However, when it comes to practical deployment, it might be challenging to execute decision making due to lack of relevant information. To give an example, implementation of a distributed AoI-based MAC protocol implies remote tracking of AoI within the data link layer of the source, although AoI is an application layer metric defined at the receiver. Furthermore, to identify the value of information within lower layers, certain knowledge about its content, context and communication purpose is crucial. This comes with the introduction of various new metrics and requires system-wide flexibility and programmability of the communication stack.

The system research is currently lacking behind theory due to the limited availability of easily programmable platforms. Additionally, the increased complexity and design challenges that are hidden prior to deployment constitute a barrier to practical implementation .
With this work, we aim to encourage researchers towards a tighter integration of practice into theory and vice versa. Moreover, we intend to provide initial design considerations and insights into customized protocol implementation for AoI and NCSs communities.

\bibliography{bibliography}
\bibliographystyle{IEEEtran}

\end{document}